\documentclass[aps, prd, onecolumn, tightenlines, notitlepage, superscriptaddress, nofootinbib, preprintnumbers, floatfix,showkeys,11pt]{revtex4-2}

\usepackage[normalem]{ulem}
\usepackage{amssymb}
\usepackage{amsmath}
\usepackage{graphicx}
\usepackage{url}
\usepackage{ulem}
\usepackage[utf8]{inputenc}
\usepackage{comment}
\usepackage[T1]{fontenc}
\usepackage{appendix}
\usepackage{titlesec}

\usepackage[x11names]{xcolor}
\usepackage{xspace}

\usepackage[export]{adjustbox} % Center figures inside equations
\usepackage{textgreek} % To add greek letters to the text
\usepackage{caption, subcaption} % Better figure customization
\usepackage{ragged2e} % for the \justifying macro
\DeclareCaptionJustification{justified}{\justifying}
\usepackage{booktabs} % Cool tables
\usepackage{orcidlink}

\usepackage{hyperref}
\usepackage[capitalise]{cleveref}
\hypersetup{colorlinks,citecolor= nicered,linkcolor= blue}
\definecolor{nicered}{rgb}{0.7,0.1,0.1}
\definecolor{nicegreen}{rgb}{0.1,0.5,0.1}

\definecolor{plotblue}{HTML}{1e88e5}
\definecolor{plotorange}{HTML}{fca009}
\definecolor{plotred}{HTML}{d81b60}
\definecolor{plotgreen}{HTML}{2c6e49}

\newcommand{\dd}{\mathrm{d}}

\definecolor{byzantium}{rgb}{0.44, 0.16, 0.39}

\makeatletter
    \newcommand{\colorboxed}[3][white]{\fcolorbox{#2}{#1}{\m@th$\displaystyle#3$}}
\makeatother

\AtBeginDocument{\hypersetup{citecolor=byzantium,linkcolor=byzantium,urlcolor=byzantium}}
\usepackage{appendix}

\definecolor{mediumpersianblue}{rgb}{0.0, 0.4, 0.65}

\definecolor{amber}{rgb}{1.0, 0.49, 0.0}

\newcommand{\rvec}{\mathbf{r}}
\newcommand{\xvec}{\mathbf{x}}

\newcommand{\rpvec}{\mathbf{r}^\prime}

\newcommand{\tauvec}{\boldsymbol{\tau}}

\newcommand{\hPBH}{hPBH\xspace}
\newcommand{\lPBH}{$\ell$PBH}
\newcommand{\lPBHs}{$\ell$PBHs\xspace}
\newcommand{\hPBHs}{hPBHs}

\newcommand{\rta}{\ensuremath{ r_\mathrm{ta}}}
\newcommand{\Msun}{M_\odot}
\newcommand{\rS}{r_\mathrm{S}}

\begin{document}

\title{{\Large The swallowed spike: the formation of light primordial black hole  structures around heavy seeds}}

\author{Agnese Tolino~\orcidlink{0009-0003-3278-0902}}
\email{atolino@ific.uv.es}
\affiliation{Instituto de F\'{i}sica Corpuscular (IFIC), CSIC‐Universitat de Val\'encia, E-46980 Valencia, Spain}

\author{Francesca Scarcella~\orcidlink{0000-0003-2456-9501}}
\email{scarcella@ifca.es}
\affiliation{Instituto de F\'{i}sica de Cantabria (IFCA, UC-CSIC), Av. de Los Castros s/n, 39005 Santander, Spain}

\author{\\Bradley J. Kavanagh~\orcidlink{0000-0002-3634-4679}}
\email{kavanagh@ifca.es}
\affiliation{Instituto de F\'{i}sica de Cantabria (IFCA, UC-CSIC), Av. de Los Castros s/n, 39005 Santander, Spain}

\author{Valentina De Romeri~\orcidlink{0000-0003-3585-7437}}
\email{deromeri@ific.uv.es}
\affiliation{Instituto de F\'{i}sica Corpuscular (IFIC), CSIC‐Universitat de Val\'encia, E-46980 Valencia, Spain}

\author{Daniele Gaggero~\orcidlink{0000-0003-3534-1406}}
\email{daniele.gaggero@pi.infn.it
}
\affiliation{Dipartimento di Fisica E. Fermi, Università di Pisa, Largo B. Pontecorvo 3, I-56127 Pisa,
Italy}
\affiliation{INFN, Sezione di Pisa, Largo Bruno Pontecorvo 3, I-56127 Pisa, Italy}

\begin{abstract}
Spikes are steep enhancements in the dark matter (DM) distribution around a heavy compact object. If the compact object is primordial, and the bulk of the DM is also composed of (lighter) primordial compact objects, for instance asteroid-mass primordial black holes (PBHs), the phenomenology of spike formation is highly non-trivial. In fact, lighter PBHs have negligible angular momentum at formation with respect to the massive central object and would therefore be captured unless enough torque is exerted from either small-scale or large-scale matter fluctuations. In this paper, we present the first comprehensive assessment of this scenario. We define the mechanisms and the initial conditions that allow light PBHs to avoid capture. We then quantify the different types of torque and follow the corresponding angular momentum evolution with a combination of analytical prescriptions and numerical simulations. We find that in the innermost region no mechanism studied here is capable of providing sufficient torque; the resulting inner core is expected to be significantly less dense than in particle scenarios, with potential consequences for the spike phenomenology and its observational signatures.
\end{abstract}

\maketitle
\tableofcontents
\section{Introduction}
\label{sec:intro}

Primordial Black Holes (PBHs)~\cite{Zeldovich:1967lct,Hawking:1971ei,1975ApJ...201....1C} are hypothetical compact objects which may have formed in the early Universe from the collapse of large-amplitude overdensities. 
Such overdensities may arise from non-trivial inflationary dynamics or from other early-Universe mechanisms, including first-order phase transitions and bubble collisions~\cite{Carr:2020gox, Green:2020jor,  Escriva:2022duf,Carr:2026hot}.
Over the years, PBHs have been widely discussed in connection with a range of observational hints and outstanding puzzles in cosmology and astrophysics, including the detection of microlensing events towards the Galactic bulge and the Large Magellanic Cloud \cite{Niikura:2019kqi,2020A&A...636A..20W, Key:2026gmp, Key:2026wbg,Udalski:2026gik}; the existence of super-early galaxies at $z > 10$; the formation of supermassive black holes \cite{Carniani:2024,Perez-Gonzalez:2025bqr,Matteri:2025vnv}; and, finally, the properties of the black-hole binaries detected by the LIGO-Virgo-KAGRA (LVK) collaboration~\cite{KAGRA:2021duu,LIGOScientific:2026wfs}.
In addition,
they constitute a viable dark matter (DM) candidate, provided they have not completely evaporated by now via Hawking radiation~\cite{Carr:2020gox, Carr:2021bzv}, with mass-dependent constraints from a variety of astrophysical and cosmological observations~\cite{Montero-Camacho:2019jte,PBHbounds,Carr:2026hot}.
The detection of one or more such objects, even if they were to constitute only a subdominant component of DM, would provide valuable insight into the elusive nature of DM~\cite{Bertone:2019vsk} and early-Universe physics.

PBHs can form over a very broad range of masses, depending on the size of the primordial overdensities they originate from. 
Indeed, realistic formation scenarios generally predict extended mass distributions, even when PBHs originate from enhanced primordial fluctuations peaked at a specific scale~\cite{Carr:2017jsz,PhysRevLett.70.9}. 
On the other hand, preferred mass scales exist, connected to the evolution of the equation of state of the primordial fluid: a lower density threshold for PBH formation is predicted in correspondence with variations of the particle content of the fluid, the most prominent of which happens at the QCD phase transition~\cite{Jedamzik:1998hc,Byrnes:2018clq, Carr:2019kxo}. This transition facilitates the formation of PBHs around $1 \, \Msun$. 
The strongest observational constraints at this mass come from the OGLE microlensing survey and LVK observations, excluding that they constitute more than 1\% of the DM~\cite{PBHbounds,Mroz:2024wag,Andres-Carcasona:2026avd}.

Heavier PBHs, despite being more strongly constrained, are of particular interest, since they could act as seeds for the formation of supermassive black holes, whose observed abundance at high redshift remains difficult to reconcile with standard black-hole and galaxy formation scenarios~\cite{Ferrara:2022dqw}. Recently, the James Webb Space Telescope observation of ``Little Red Dots'' has further motivated the idea that heavy seeds may have existed at early times \cite{Pacucci:2023oci}. A subdominant fraction of PBHs is sufficient to provide such seeds in the required abundance. 
Cosmic microwave background (CMB) spectral distortions and anisotropy constraints strongly limit PBHs above $\sim 10^3$--$10^4 M_\odot$~\cite{Karam:2022nym,Byrnes:2024vjt} setting an upper limit on the mass of viable primordial seeds~\cite{DeLuca:2025nao}.
At this mass, PBHs are constrained to constitute no more than a fraction $\sim 10^{-4}$ of the DM~\cite{Agius:2024ecw}. 
Finally, PBHs remain a viable candidate to constitute the entirety of the DM within a narrow mass window, covering asteroid-mass  objects, approximately from $10^{-16}\,M_\odot$ to $ \sim 10^{-11}\,M_\odot$~\cite{Montero-Camacho:2019jte,Tinyakov:2024mcy}.

Motivated by the above considerations, we focus on scenarios in which the bulk of the DM is composed  of ``light'' PBHs (\lPBHs) around the asteroid mass range,
accompanied by a subdominant population of heavier PBHs (\hPBHs) with masses around $1 \, M_\odot$ or above (see, for instance, the models discussed in~\cite{Carr:2019kxo,Franciolini:2022tfm}).
The coexistence of these two populations would lead to the  early formation of gravitationally bound structures, populated by light PBHs and seeded by the heavy ones. 

Such configurations have attracted interest in the literature due to their possible observational signatures, including an enhancement of gravitational-wave emission in their high-density environments~\cite{Nishikawa:2017chy,Kuhnel:2018mlr,Feng:2024obn}. However, comparatively little attention has been devoted to the formation process and resulting density profiles of these structures, and to how these differ from the analogous structures formed in particle DM scenarios. 

Indeed, so-called DM \textit{spikes} around PBHs have been extensively studied in the context of weakly interacting massive particles (WIMPs) (see, e.g., Refs.~\cite{Mack:2006gz,Lacki:2010zf,Eroshenko:2016yve,Adamek:2019gns,Carr:2020mqm}). 
If these constitute the bulk of the DM, while coexisting with PBHs, extremely dense WIMP structures are expected to form around each PBH. 
These structures would form as WIMPs decouple from the background expansion, turning around under the gravitational pull of the compact object and becoming gravitationally bound to it. 
The formation process begins immediately after the collapse of the fluctuation originating the PBH, typically during radiation domination, and proceeds in shells. Particles at increasing comoving distances from the PBH reach the turn-around radius $\rta$ at later turn-around times $t_\mathrm{ta}$, with $ \rta^3 \simeq 2 G M t_\mathrm{ta}^2 \,$\cite{Adamek:2019gns}, where $M$ is the PBH mass. Assuming that, up to the time of turn-around, the DM density dilutes with the Hubble expansion, one obtains the following relation (valid in the radiation era) between the turn-around radius and the DM shell density $\rho$,
\begin{equation}
\label{eq:rho_ta}
    \rho(\rta) \simeq  \rho_\mathrm{eq}^\mathrm{DM} \,  (2 G M )^{3/4} \,  t_\mathrm{eq}^{3/2} \, r_\mathrm{ta}^{-9/4} \, ,
\end{equation}
displaying a steep $r^{-9/4}$ scaling ($t_\mathrm{eq}, \rho_\mathrm{eq}^\mathrm{DM}$ are respectively the age and DM density of the Universe at matter-radiation equality). 
The physical picture leading to \cref{eq:rho_ta} can be directly applied to scenarios in which the bulk of the DM consists of \lPBHs.

However, \cref{eq:rho_ta} does not yet describe the spike density profile, unless one assumes that shells ``freeze'' at the time of decoupling. Instead, DM particles continue their infalling motion and thereafter oscillate in elliptic orbits around the PBH. 
Assuming pairwise DM interactions to be negligible, the distribution of orbital parameters can be obtained from that of the DM peculiar velocities at decoupling. Then, integrating the time spent by each particle at a given radius, one finally obtains the radial density profile of the DM spike~\cite{Eroshenko:2016yve,Adamek:2019gns,Carr:2020mqm}.
For WIMPs, expected to be thermally coupled at early times, the velocity distribution at turn-around is thermal.
The resulting radial density profile displays a $\rho \propto r^{-9/4}$ scaling for a large portion of  parameter space,
but regimes with shallower slopes also occur~\cite{Carr:2020mqm}. The crucial point here is that each shell described by \cref{eq:rho_ta} reaches its stable dynamic configuration within a free-fall time, and this configuration is determined by the initial conditions (density and velocity dispersion) at turn-around.

In the case of \lPBHs, it is not possible to define a thermal distribution of velocities at the time of decoupling, meaning the profile calculation of Refs.~\cite{Eroshenko:2016yve,Adamek:2019gns,Carr:2020mqm} cannot be directly applied.  In fact, PBHs are expected to form with negligible peculiar velocities with respect to the background expansion~\cite{Carr:1974nx, Nakamura1997, 
Ali-Haimoud:2017rtz, Raidal:2017mfl}. A vanishing kinetic energy at the time of turn-around would naively lead to radial infall, with the \lPBHs being accreted by the \hPBH. As a consequence, {\it a natural question that arises is whether an \lPBH~spike can ever form at all.}

\lPBHs can avoid being accreted if torques, acting during infall, provide them with sufficient angular momentum. A very similar situation arises in the formation of PBH binaries, where the torques necessary to avoid a head-on collision between the pair are induced by other PBHs (and adiabatic DM perturbations) surrounding the binary~\cite{Nakamura1997,Eroshenko:2016hmn,Ali-Haimoud:2017rtz,Raidal:2018bbj}.
The goal of the present work is to study the physical mechanisms that are capable of producing torques on the infalling \lPBH\ shells, and assess whether they can impart sufficient angular momentum to allow a spike to form. 

We identify two main classes of effects capable of generating angular momentum during the collapse.
The first class consists of torques sourced by gravitational perturbations at large distances. Distant objects or large-scale density perturbations generate a tidal field across the \lPBH-\hPBH separation. This is analogous to the mechanism that prevents head-on collisions in PBH binary formation.
The second class of effects is intrinsically local. Since the infalling shells are made of a finite number of \lPBHs, Poisson fluctuations are expected to generate small-scale anisotropies within each shell, and then we expect the effect to be twofold: {\it (i)} neighboring \lPBHs exert non-vanishing torques on each other with respect to the central \hPBH; {\it (ii)} the same anisotropy exerts a net force on the \hPBH, causing it to acquire angular momentum with respect to the \lPBHs. 

For each of these mechanisms, our goal is to compute the accumulated angular momentum  of a \lPBH\ turning around at radius $r_{\rm ta}$, and to compare it with the capture threshold. This comparison defines, for each mechanism, a critical turn-around radius $r_{\rm c}$. Shells with $r_\mathrm{ta} < r_{\rm c}$ remain in the loss cone and are swallowed by the central object, whereas shells with $r_\mathrm{ta} > r_{\rm c}$ acquire enough angular momentum to survive and contribute to the density enhancement. The main result of the paper is therefore a map of this survival criterion, for each source of torque, in the $(m, r_\mathrm{ta})$ plane, $m$ being the \lPBH~mass. 

The paper is structured as follows. In \cref{sec:setup} we introduce our assumptions and physical setup. In \cref{sec:external_torques} we neglect pairwise interactions between \lPBHs and study the effect of tidal fields generated by other heavy PBHs and  adiabatic density fluctuations on large scales. In~\cref{sec:ell-ell} we consider angular momenta arising from the interactions of each \lPBH~with its neighbors. In \cref{sec:shell_capture} we study torques associated to the force induced on the \hPBH by the anisotropy of \lPBH\ shells. In Sec.~\ref{sec:results} we present numerical results for each mechanism, in terms of the minimum turn-around radius associated to the survival of an \lPBH\ shell. Finally, in Sec.~\ref{sec:conclusions}, we discuss our conclusions.

\section{Physical setup}
\label{sec:setup}

Throughout this work, we consider a single mass $m \ll M$ for the \lPBH~population, spanning a broad range starting at $10^{-16} M_\odot$ and up to  $10^{-5} M_\odot$. For $M$, we consider two benchmark values: $1 \,\Msun$, motivated by the QCD peak, and $10^3 \Msun$, for the highest viable initial mass for supermassive black hole seeds. We assume both populations to be Poisson distributed at formation. 
Unless otherwise specified, we assume \lPBHs~to constitute the majority of DM, $f_\ell \equiv \Omega_\mathrm{\ell PBH}/\Omega_\mathrm{DM} \simeq 1$ and the heavy PBHs to represent a subdominant DM fraction, $f_h \equiv \Omega_\mathrm{h PBH}/{\Omega_\mathrm{DM}} \ll 1$. 
Note that both heavy and light PBHs are expected to form shortly after the end of the inflation and before Big Bang nucleosynthesis (BBN) for the whole range of masses considered here ($t_{\rm formation} \sim 10^{-21} \,{\rm s}$ for the asteroid-mass scale, and $t_{\rm formation} \sim 10^{-6} \,{\rm s}$ for the solar-mass scale). These timescales are shorter than any other relevant timescale considered in this work.

As discussed in the introduction, the evolution of \lPBHs around a heavy seed mirrors that of WIMPs up to the time of decoupling from the Hubble flow. Hence, the turn-around density $\rho(r_\mathrm{ta})$ of each shell is given by \cref{eq:rho_ta} and its number density is
\begin{equation}
\label{eq:n_ta}
    \bar n_\ell (\rta) \simeq  \dfrac{f_\mathrm{\ell} \,\rho(r_\mathrm{ta}) }{m} \, .
\end{equation}
\lPBHs~are expected to form with negligible peculiar velocities with 
respect to the Hubble flow, and we now argue that they remain approximately comoving up to the time of turn-around.

Peculiar velocities build up as a consequence of \lPBH\ structure formation, a process which proceeds bottom-up starting with the formation of binaries. The typical redshift associated to this process thus provides the reference time scale for the build-up of peculiar velocities.
An \lPBH\ pair at initial comoving separation $x$ decouples from the background expansion at  $a_\mathrm{dec} \approx a_\mathrm{eq} (x/\bar x)^3/f_\ell$~\cite{Nakamura1997}, where $\bar x$ is the mean nearest-neighbor separation and $a_\mathrm{eq}$ is the scale factor at matter-radiation equality. For typical values, $ x \sim \bar x$, decoupling occurs around matter-radiation equality for $f_\ell\sim1$, and later for smaller PBH fractions\footnote{Binaries expected to merge within the age of the Universe correspond to rare configurations with $x \ll \bar{x}$; these decouple deep in the radiation era.}. 
Since we restrict our analysis to shells that decouple in the radiation era, we expect peculiar velocities to remain negligible at turn-around for most \lPBHs in the shell. 
By the same argument, the spatial distribution of \lPBHs at turn-around can be assumed to not have departed significantly from a Poissonian. 

Based on these considerations, we assume that at the time of decoupling from the Hubble flow, \lPBHs have null peculiar velocities and are Poisson distributed with mean separation determined by the shell density
\begin{align}
\label{eq:light_separation}
    \bar{x}(r_\mathrm{ta}) &\simeq \left( \dfrac{3}{4 \pi  \bar n_\ell (\rta)}\right)^{1/3} \, .
\end{align}
We define the spike radius $r_\mathrm{sp, eq}$ as the turn-around radius of the shell that decouples at matter-radiation equality
\begin{equation}
\label{eq:r_sp_eq}
    r_\mathrm{sp, eq} \simeq \left(2GM t_{\rm eq}^2\right)^{1/3} \approx 3.4 \times 10^{-2} \, \mathrm{pc} \, \left( \dfrac{M}{M_\odot}\right)^{1/3} \,
    \approx 3.5 \times 10^{11} \,r_\mathrm{S} \left( \dfrac{M}{M_\odot}\right)^{-2/3}  \, ,
\end{equation}
where $\rS$ is the Schwarzschild radius of the \hPBH.
This radius encloses the central core of a typical DM spike around a PBH and, because of its high density, represents its most phenomenologically relevant part. 
Shells at $r > r_{\rm sp, eq}$ continue to be accreted during matter domination; however, our assumptions of negligible peculiar velocities and Poissonian statistics at turn-around no longer hold beyond $r_{\rm sp, eq}$. 

After turn-around, shells start infalling, while the enclosed \lPBHs~acquire angular momentum with respect to the central \hPBH, through the torques discussed in~\cref{sec:external_torques,sec:internal_torques}. 
We treat both the infall and the action of torques in Newtonian gravity. Relativistic corrections are expected to become important only near the Schwarzschild radius of the \hPBH, but as we justify below, this late phase does not contribute significantly to the accumulated angular momentum. 

To leading order, a torque $\tau$ acting throughout the infall from 
$r_{\rm ta}$ imparts a specific angular momentum $\ell \approx 
t_{\rm ff}(r_{\rm ta})\,\tau/m$, where $t_{\rm ff}(r) = \pi/(2\sqrt{2})
r^{3/2}(GM)^{-1/2}$ is the free-fall time. More precisely,
\begin{equation}
\label{eq:ell_of_tau}
    \ell(\rta) =
    \int_{0}^{t_\mathrm{ff}} 
    \frac{\tau(t)}{m}\,\mathrm{d}t\, \simeq \lim_{r_\mathrm{f} \to 0} \int_{\rta}^{r_\mathrm{f}} %\frac{\tau(t)}
    \frac{\tau(r)}
    {m} \dfrac{1}{v_r(r)}\,\mathrm{d}r\,, 
\end{equation}
where the radial infall velocity in the potential of the \hPBH~is
\begin{equation}
\label{eq:vel_infall}
v_r(r) = -\sqrt{2GM\left(\frac{1}{r}-\frac{1}{r_{\rm ta}}\right)},
\end{equation}
with the sign chosen such that $v_r < 0$ during infall.

In writing \cref{eq:ell_of_tau}, we have assumed the trajectory remains close to 
radial.
This approximation may be expected to break down at small radii; however, most of the angular momentum is accumulated at large radii, where the infall velocity is small. More precisely, evaluating the integral in \cref{eq:ell_of_tau} for a fixed torque shows that $90\%$ of the angular momentum is accumulated at $r \gtrsim 0.35 \, \rta$. Our goal is to identify the threshold turn-around radius for which the \lPBHs narrowly escape the Schwarzschild radius $r_\mathrm{S}$ of the \hPBH; as long as this threshold radius is much larger than $r_\mathrm{S}$, it is safe to assume that the trajectory is very close to radial for its relevant portion\footnote{This condition is verified a posteriori in \cref{sec:results}.}.

Having estimated the acquired angular momentum, we compare it to the minimum value $\ell_\mathrm{min} $ necessary to avoid a head-on merger. Including relativistic corrections, $\ell_\mathrm{min} \simeq 4 (G/c) M $~\cite{Bertone:2024wbn}, where $M$ is the mass of the hPBH. 
Since $\ell(r_{\rm ta})$ depends on the turn-around radius, the condition 
$\ell(r_{\rm ta}) > \ell_{\rm min}$ defines a critical radius $r_c$: 
shells with $r_{\rm ta} < r_c$ are expected to be accreted by the 
central \hPBH, while those at $r_{\rm ta} > r_c$ survive and contribute 
to the spike. In the following sections we identify the dominant sources 
of torque and derive the corresponding critical radius for each.

\section{Torques from external sources}
\label{sec:external_torques}

We begin by considering the torques generated by gravitational sources at large distance from the system. In this section, we make the assumption that interactions between \lPBHs are negligible and that the \hPBH~can be considered at rest (we will return on these assumptions in \cref{sec:ell-ell,sec:shell_capture}, showing that they do not always hold). Under these approximations, the infall of each \lPBH~towards the \hPBH~can be treated as the formation of an independent, high mass-ratio binary. 
Then, we can directly apply the treatment of Ref.~\cite{Nakamura1997, Ali-Haimoud:2017rtz}, studying the angular momentum induced by other \hPBHs~and by adiabatic large-scale perturbations in the DM density field. 

As long as the distance to the external sources remains much larger than the \lPBH–\hPBH separation $r$, the induced gravitational field varies smoothly across the pair and can be described through a tidal expansion.
The tidal tensor is defined as ${T}_{ij}=\partial_i\partial_j\Phi_{\rm ext}$, where $\Phi_{\rm ext}$ is the gravitational potential. Expanding around the center of mass of the pair, the differential acceleration to leading order in $r/R$ (with $R$ the typical distance to the external sources) is  $\bold{a} = \bold{T}
\cdot \Delta \bold r$, 
where $\Delta \bold r$ is the displacement of the \lPBH~from the 
center of mass.
We neglect the displacement of the \hPBH~from the center of mass and approximate $\Delta \bold r \approx \bold r$.
The induced torque depends on the tidal tensor through the cross product of the pair separation
\begin{align}
\label{eq:tid_torque_general}
 \dfrac{\tauvec}{m}
 &= \bold{r} \times \left( \bold{T}\cdot \bold r \right)\, .
\end{align}
from which the accumulated angular momentum follows via~\cref{eq:ell_of_tau}.

\subsection{Torques from other heavy PBHs}
\label{sec:hPBH_ext}

\begin{figure}
    \centering
    \includegraphics[width=0.8\linewidth]{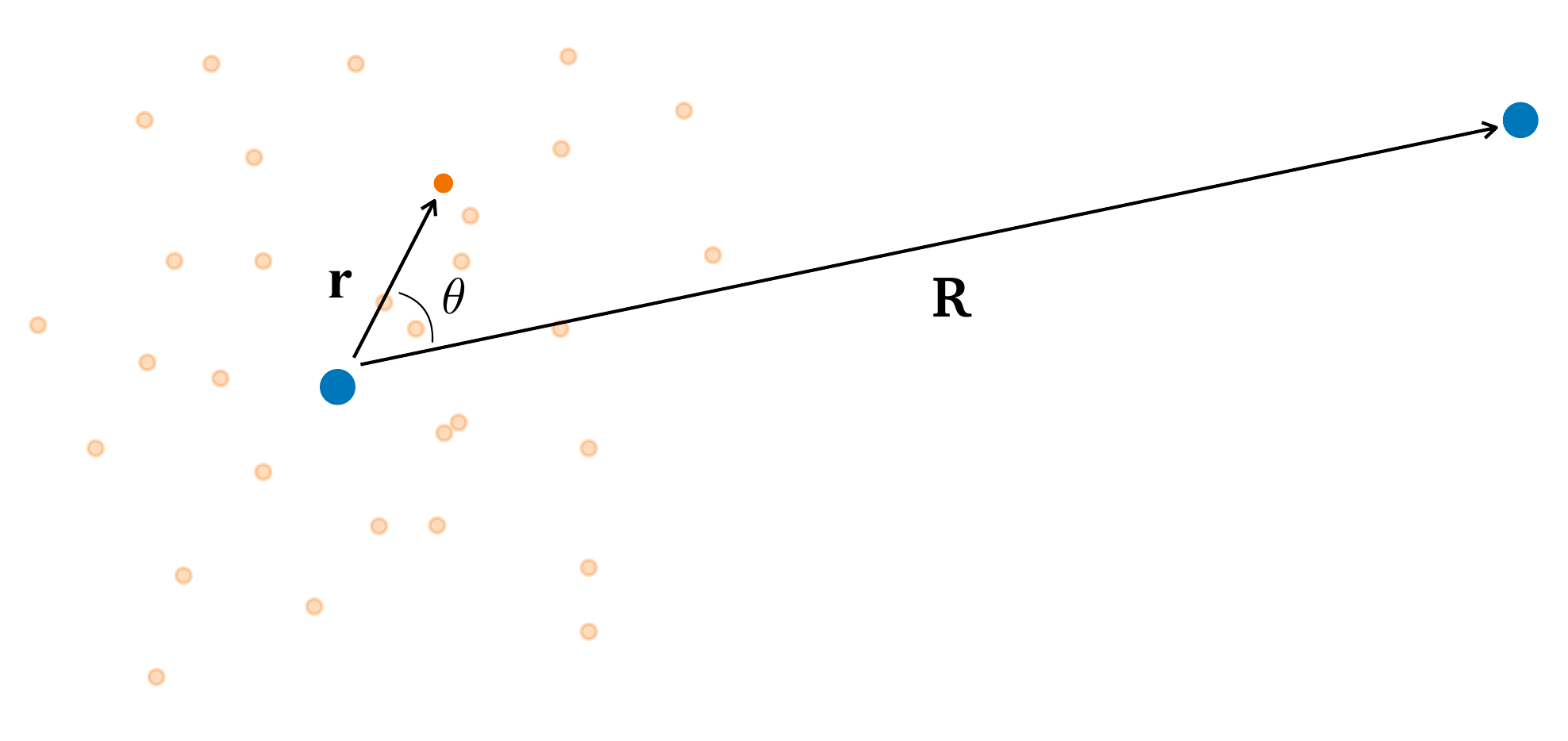}
    \caption{Schematic of the setup for the estimate of torques from \hPBH~neighbors (blue dots). We treat each light-heavy pair independently, neglecting interactions with the other \lPBHs (orange). The torque is dominated by the nearest neighbor, placed at distance $\mathbf R$ from the center of mass of the pair. $\mathbf r$ is the light-heavy separation and $\theta$ represents the angle between $\mathbf R$ and $\mathbf r$.}
    \label{fig:hPBH_torques}
\end{figure}

We first consider torques generated by neighboring \hPBHs, assumed to be Poisson distributed with mean separation
\begin{equation}
\label{eq:heavy_separation}
\bar{R}(a) \simeq \dfrac{a}{a_\mathrm{eq}}\left( \dfrac{3M}{4 \pi f_h \rho_\mathrm{eq}^\mathrm{DM}}\right)^{1/3} , 
\end{equation}
where $f_h \ll 1$ is the \hPBH~fraction of DM. Comparing this separation to the scale factor-dependent turn-around radius, we find $\rta/\bar R \; \lesssim \; 0.5 \, f_h^{1/3}$ for $a \,<\, a_\mathrm{eq}$. Hence, for $f_h \ll 1$, we are always in the tidal regime $R \gg r$.
The tidal tensor sourced by a single \hPBH~at distance $R \gg r$ (see Fig.~\ref{fig:hPBH_torques}) is
\begin{equation} {T}_{ij}= -\frac{G M}{R^{3}}\left( \delta_{ij}-3\hat R_i\hat R_j\right) \, , \end{equation}
and the induced torque on the \lPBH~is
\begin{align}
\label{eq:torque_hPBH}
\tauvec
&=  \,\dfrac{3GM m r^2}{R^3} \, (\hat{\bold{r}} \times \hat{\bold{R}} ) \, (\hat{ \bold{r}} \cdot\hat{\bold{R}} ) \,,
\end{align}
with magnitude
\begin{align} 
\tau &= \frac{3}{2} \dfrac{GM m r^2}{R^3} \,|\sin2\theta | \, ,
\end{align}
where $\theta$ is the angle between $\mathbf R$ and $\mathbf r$. Since the \hPBHs~are Poisson distributed, the torque direction is statistically isotropic in the plane orthogonal to $\rvec$. 
Averaging $\tau$ over the orientation of $\mathbf R$ and setting $\bar R$ to the typical separation (\cref{eq:heavy_separation}) yields the characteristic torque from the nearest neighbor
\begin{equation}
\label{eq:tau_c_hPBH}
 \tau_\mathrm{c} (r)= \dfrac{GMm r^2}{\bar {R}^3} \,.
\end{equation}

We now consider the total torque, given by the sum of contributions from all other \hPBHs; this is given by
\begin{align}
\label{eq:torque_hPBH_tot}
\tauvec
&=  \,3GM m \,r^2 \sum_k \dfrac{\, (\hat{ \mathbf{r}} \times \hat{\mathbf{R}}_k ) \, (\hat{ \mathbf{r}} \cdot\hat{ \mathbf{R}}_k ) }{{R_k^3} }\,.
\end{align}
We determine its probability distribution  integrating over all possible configurations of the \hPBHs, following the prescriptions in \cite{Ali-Haimoud:2017rtz} and \cite{Chandrasekhar:1943ws} (see \cref{app:subsec_Markov}). The total torque on a given \lPBH~at position $r$ has magnitude $\tau= \beta \tau_\mathrm{c}$, where $ \tau_\mathrm{c}$ is given by \cref{eq:tau_c_hPBH} and $\beta$ follows the distribution 
\begin{equation}
    \mathcal{P}(\beta) = \dfrac{\beta}{\left( 1+\beta^2\right)^{3/2}} \,.
\end{equation}
We integrate the specific torque over the free-fall path (\cref{eq:ell_of_tau}), neglecting the growth of $\bar R$ during the free-fall time, and obtain the angular momentum
\begin{align}
\ell
= \beta \, \frac{ 5\pi}{16} \frac{\left(GM\right)^{1/2}}{\overline{R}^3}\, r_\mathrm{ta}^{7/2} \equiv \beta \ell_{\rm c}\,,
\end{align}
where $\ell_c$ is the characteristic angular momentum.
To estimate the critical radius for \lPBH~capture, we compare the characteristic angular momentum $\ell_{\rm c}$ to the escape threshold $\ell_{\rm min}$, obtaining
 \begin{align}
    r_\mathrm{c} 
     & \approx 5.43 \times  10^{-3} \; \mathrm{pc} \, \left( \dfrac{M}{M_\odot}\right)^{3/5} \left( \dfrac{f_h}{10^{-4}}\right)^{-4/5}  \nonumber \\
    &\approx  0.16 \,  r_\mathrm{sp, eq} \, \left( \dfrac{M}{M_\odot}\right)^{4/15}\left( \dfrac{f_h}{10^{-4}}\right)^{-4/5}  \, .
\end{align}
We find that external \hPBHs~can therefore generate sufficient angular momentum 
to prevent accretion for $M \lesssim M_\odot$, while the critical radius exceeds the turn-around radius at equality for 
$M \gtrsim 10^3\,M_\odot$ and $f_h \approx 10^{-4}$.

\subsection{Torques from large-scale cold dark matter density perturbations}
\label{sec:adiabatic}

For small \hPBH fractions, we expect the tidal field to be dominated by the adiabatic component of large scale density fluctuations, whether in the form of particle DM or \lPBHs\cite{Eroshenko:2016hmn, Ali-Haimoud:2017rtz}. 
The linear density contrast $\delta_{\rm cdm}\equiv \Delta \rho/ \bar \rho$ satisfies the Poisson equation in physical coordinates $\nabla^2 \delta \Phi = 4\pi G\, \bar{\rho}\, \delta_{\rm cdm}\,,$ where $\bar{\rho}$ is the background cold dark matter (CDM) density.

First, we compute the variance of the tidal torque (\cref{eq:tid_torque_general}), which is given by 
\begin{equation}
\label{eq:tau_cdm_tensor}
    \dfrac{{\langle\tau^2\rangle}}{m^2} = r^4 \langle  \epsilon_{ijk} {\hat r}_j T_{kl} {\hat r}_l \,\epsilon_{ipq} {\hat r}_p T_{qm} {\hat r}_m \rangle = r^4 \left( \frac{1}{5} \langle T_{ij} T_{ij} \rangle - \frac{1}{15} \langle T_{ii} T_{jj}\rangle\right) \, ,
\end{equation}
where in the last equality we have taken the average over the orientation of $\hat{\mathbf{r}}$, assumed isotropically distributed. The angular brackets $\langle \ldots \rangle$ correspond to the average over realizations of the matter field. The tidal tensor can be obtained in Fourier space from the Poisson equation
\begin{equation}
    \tilde{T}_{ij} (k) =  4 \pi G \, \bar\rho_\mathrm{eq}  \dfrac{k_i k_j}{k^2} \, \delta_{\rm cdm} (k) \, .
\end{equation}
Using the definition of the power spectrum $ \langle  \delta_{\rm cdm} (k)  \delta_{\rm cdm} (k') \rangle  = (2 \pi)^3 \delta (k-k') \mathcal{P}_{\rm cdm}(k)$, we have
\begin{align}
\label{eq:tid_tensor_delta}
    \langle T_{ij} T_{ij}\rangle = \langle T_{ii} T_{jj}\rangle 
    &=  \left( 4 \pi G \, \bar\rho \right)^2 \int \dfrac{d k}{2 \pi^2} k^2 \mathcal{P}_{\rm cdm}(k)  \, \equiv  \left( 4 \pi G \, \bar\rho \right)^2 \sigma^2  \,, 
\end{align}
where we have identified $\sigma^2= \langle\delta_\mathrm{cdm}^2 \rangle$, the variance of the CDM power spectrum to be evaluated on the relevant scales, between the \lPBH-\hPBH~separation and the horizon scale. Substituting this expression in \cref{eq:tau_cdm_tensor}, we find the characteristic torque 
\begin{align}
     \tau_\mathrm{c}
    &=  \sqrt{ \frac{32 }{15} } \pi \,  G m \, r^2\,  \bar \rho \,
    %\big|_{a_\mathrm{ta}}
    \sigma    \, .
\end{align}
Keeping $\bar \rho$ and $\sigma$ constant\footnote{We have checked that including the $a^{-3}$ scaling of $\bar \rho$ reduces the torque by a factor 2. We neglect the logarithmic growth of perturbations during radiation era.}, we integrate along the free-fall path (\cref{eq:ell_of_tau}), obtaining the specific angular momentum
\begin{align}
\label{eq:ell_CDM}
    \ell_{\mathrm c}
    &= \sqrt{\frac{5 }{48}} \pi^2 \sqrt{\frac{G}{M}} \bar \rho \,
    \sigma \,\rta^{7/2} \, ,
\end{align}
where $\bar \rho$ is to be evaluated at turn-around.

\begin{figure}
    \centering
    \includegraphics[width=\linewidth]{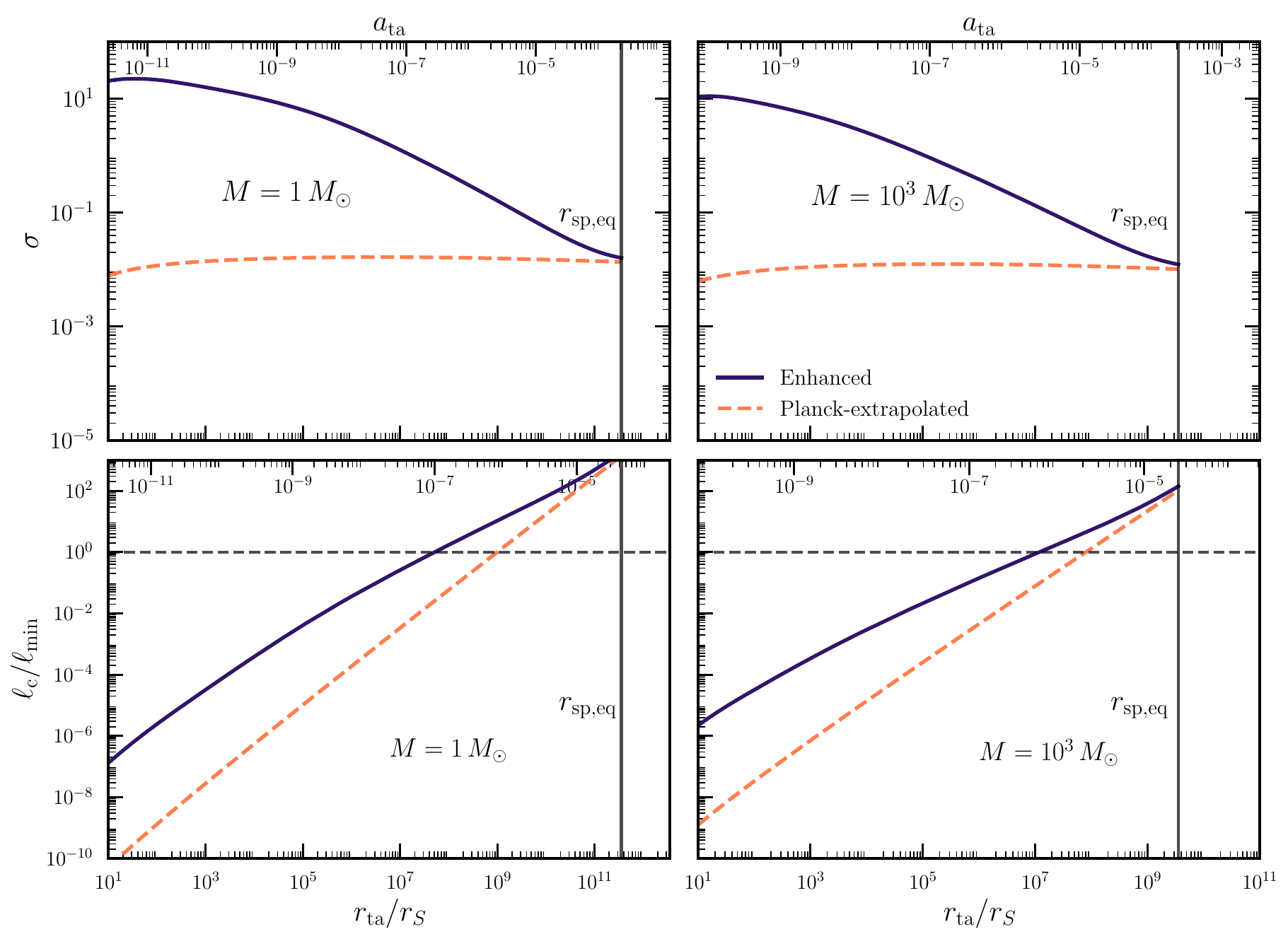}
    \caption{\textbf{Top panels}: Variance $\sigma$ of primordial CDM fluctuations, as a function of the turn-around radius, normalized to the Schwarzschild radius of the \hPBH, for $M_\odot$ (left) and $10^3 M_\odot$ (right). \textbf{Bottom panels}: Ratio of the accumulated angular momentum $\ell_{\rm c}$ to the capture threshold $\ell_{\rm min}$, as a function of the same normalized turn-around radius. In both panels, the blue solid line corresponds to the enhanced primordial power spectrum of~\cite{Franciolini:2022pav}, shifted to peak at the corresponding \hPBH~mass, while the coral dashed line assumes a Planck-extrapolated~\cite{Planck:2018vyg} power spectrum. The horizontal dashed line in the bottom panels marks $\ell_{\rm c}/\ell_{\rm min} = 1$; shells above this threshold avoid direct capture.}
\label{fig:sigma_rta_l}
\end{figure}

To evaluate $\sigma$, we begin by following Ref.~\cite{Ali-Haimoud:2017rtz} and extrapolate the primordial power spectrum (PPS) constrained by Planck observations~\cite{Planck:2018vyg} to the small scales of interest. 
We neglect the logarithmic growth of perturbations between turn-around and equality and use the transfer functions provided by the Boltzmann solver \texttt{CAMB}~\cite{camb_notes} to obtain the power spectrum at matter-radiation equality.
We then estimate $\sigma$ numerically, integrating $\mathcal{P}_{\rm cdm}$ over the relevant scales, which are those larger than the turn-around radius; see \cref{app:subsec_CAMB} for details.
The resulting variance is shown as a dashed orange curve in the top panels of \cref{fig:sigma_rta_l}. It presents a logarithmic scaling with the turn-around radius and agrees with the value given in Ref.~\cite{Ali-Haimoud:2017rtz}.
Taking as reference $\sigma \sim \mathcal{O}( 10^{-2})$, and comparing $\ell_{\rm c}$ (\cref{eq:ell_CDM}) to the minimum value $\ell_{\rm min}$, we obtain the critical radius 
\begin{align}
\label{eq:rc_sigma}
    r_\mathrm{c}
    & \approx
    1.36 \times 10^{-4}\,\mathrm{pc}
    \left(\frac{M}{M_\odot}\right)^{3/5}
    \left(\frac{\sigma}{0.01}\right)^{-4/5}  \nonumber 
    \\
    &\approx
    4 \times 10^{-3}\,
    r_\mathrm{sp, eq}
    \left(\frac{M}{M_\odot}\right)^{4/15}
    \left(\frac{\sigma}{0.01}\right)^{-4/5} \, .
\end{align}

However, PBH formation from gravitational collapse requires  $\mathcal{O}(1)$ fluctuations on scales associated with the PBH mass. Furthermore, the growth rate of perturbations towards small scales is limited ($\mathcal{P}(k) \sim k^{4}$ in single field inflation~\cite{Byrnes:2018txb}), so that the enhancement must also affect scales larger than the \hPBH mass. Hence, extrapolating the Planck-constrained PPS is likely to significantly underestimate the value of the variance on the relevant scales.
In order to take this into account, we also evaluate $\sigma$ in a more realistic scenario, based on the enhanced PPS proposed in~\cite{Franciolini:2022pav,Franciolini:2022tfm}, constructed to produce a PBH mass function peaked at $1 \, M_\odot$ and in the asteroid mass range (for the $M = 10^3\,M_\odot$ benchmark, we shift it to obtain a peak at the corresponding mass scale; see \cref{app:subsec_PPS} for details).
The result is shown in the upper panels of \cref{fig:sigma_rta_l} as a solid blue line. The spike radius $r_{\rm sp, eq}$ is also shown for reference as a vertical black line in both panels.
For shells turning around near the Schwarzschild radius of 
the \hPBH, enhanced perturbations on small scales contribute to the tidal field and $\sigma$ is boosted by up to three orders of 
magnitude. At large turn-around radii, by contrast, smaller scales are excluded and the Planck-extrapolated value is gradually recovered.

To assess the impact of the PPS enhancement on the critical radius, we numerically compare $\ell_{\rm c}$ to $\ell_{\rm min}$ as a function of $r_{\rm ta}$; their ratio is shown in the lower panels of \cref{fig:sigma_rta_l}, with the horizontal dashed line marking 
$\ell_{\rm c} = \ell_{\rm min}$. For both the Planck-extrapolated and enhanced spectra, this ratio increases steadily with $r_{\rm ta}/r_\mathrm{S}$. In the innermost shells, the enhanced spectrum yields a value of $\ell_{\rm c}$ approximately three orders of magnitude larger than the Planck-extrapolated case, though this difference narrows at larger turn-around radii. The critical radius is ultimately set by the large-$r_{\rm ta}$ behavior of $\sigma$, where, for both spectra, $\ell_{\rm c}$ eventually exceeds $\ell_{\rm min}$. In the enhanced case, this threshold radius is about an order of magnitude smaller than in the Planck-extrapolated scenario.
Finally, we evaluate the critical radius numerically for the enhanced PPS, finding  $r_{\rm c} \simeq 10^{-4}\,r_{\rm sp, eq}$ for $M = 1\,M_\odot$ (left panel) and $r_{\rm c} \simeq 3\times10^{-2}\,r_{\rm sp, eq}$ for $M = 10^3\,M_\odot$ (right panel).

\section{Torques from internal sources}
\label{sec:internal_torques}

The torques discussed in \cref{sec:external_torques} originate from distant external sources. These generate a gravitational field that varies slowly across the light-heavy separation, hence admitting a tidal expansion. We now turn to considering rapid, small scale variations of the gravitational potential across this separation, which produce additional forces acting on each PBH independently. 

Indeed, unlike the case of PBH binary formation, the light-heavy pair infall takes place in an environment populated by other \lPBHs, which generate non-negligible Poisson-induced perturbations on small scales~\cite{Inman:2019wvr}.
We expect these local fluctuations to give rise to torques mainly through two distinct mechanisms. First, each \lPBH~is affected by the gravitational attraction of the neighboring \lPBHs. Second, the anisotropy of the \lPBH~distribution around the \hPBH can generate a net force on it, shifting it away from the center of mass of the system.
We discuss these two effects separately in the following.

\subsection{Torques from light--light PBH interactions}
\label{sec:ell-ell}

\begin{figure}[th!]
\centering
\includegraphics[clip, trim=0 0.5cm 0 0.5cm,width=0.4\textwidth]{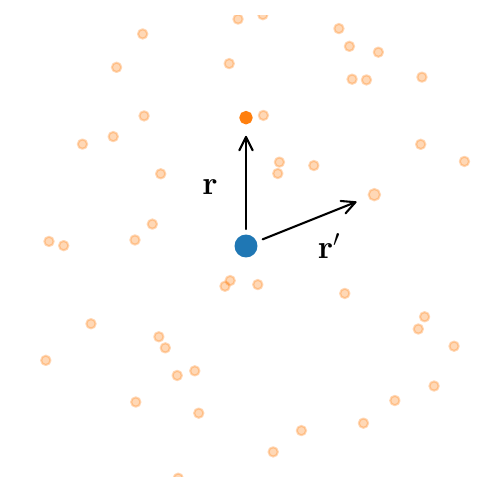}
\caption{Schematic of the setup for the estimate of torques from \lPBH~neighbors. We consider a distribution of \lPBHs of mass $m$ (orange points) around a \hPBH~of mass $M$ (blue) at the origin. We compute the torque acting on a \lPBH~at position $\rvec$ from all other \lPBHs at positions $\{\rpvec\}$.}
\label{fig:setup}
\end{figure}

In this section, we study the build-up of angular momentum through gravitational interactions among \lPBHs. 
Two limiting regimes can be distinguished depending on the number of close encounters between \lPBHs during infall. If many close encounters happen, the relative positions of the \lPBHs are reshuffled. Then, the torque on a given \lPBH~fluctuates in magnitude and direction, causing the angular momentum to build up through a stochastic process. In the opposite regime, the \lPBHs\ approximately preserve their relative spatial ordering throughout the collapse, so that the torque direction remains nearly constant. The angular momentum is then determined by the initial configuration. 
The latter regime is the one relevant to our case. Indeed, the typical nearest-neighbor crossing time is of order the free-fall time, $t_{\rm ff}$, as we show in \cref{app:orb_cross}. Consequently, only a small number of crossings occurs during the shell collapse, making topology-changing encounters rare. 

Let us first consider the torque on a \lPBH\ (at position $\mathbf{r}$ with respect to the \hPBH) generated by a single other \lPBH, located at position $\rvec^\prime$, as shown in Fig.~\ref{fig:setup}. This is given by
\begin{equation}
\tauvec = -G m^2 \frac{(\rvec - \rvec^\prime) \times \rvec^\prime}{\left| \rvec - \rvec^\prime \right|^3} = -G m^2 \frac{\rvec \times \rvec^\prime}{\left| \rvec - \rvec^\prime \right|^3} \, .
\end{equation}
The resulting initial torque will be denoted by $\tauvec_i = \tau_i \hat{\mathbf n}$, where $\hat{\mathbf n}$ is a unit vector. Since the light PBHs are uniformly distributed within each shell, there is no preferred azimuthal direction around $\rvec$. The torque direction $\hat{\mathbf n}$ is therefore statistically isotropic in the plane orthogonal to $\rvec$.  First, to have an order-of-magnitude estimate, let us assume that the two \lPBHs~are nearest neighbors, with typical separation $|\mathbf{r} - \mathbf{r}^\prime| \approx \bar{x}(r_{\rm ta})$ given by \cref{eq:light_separation}.
The magnitude of the initial torque from one \lPBH\ can hence be estimated to be
\begin{equation}
\tau_i \approx G m^2 \frac{\rta \,\bar{x} \,|\sin\vartheta|}{ 
\bar{x}^3} = \left(\frac{4\pi}{3}\right)^{2/3} \,G m^2 \overline{n}_\ell^{2/3} \rta \,|\sin\vartheta| \, ,
\end{equation}
where $\vartheta$ is the initial angle between $\rvec$ and $\rvec^\prime$. 

\begin{figure}[tb]
\centering
\includegraphics[width=0.49\textwidth]{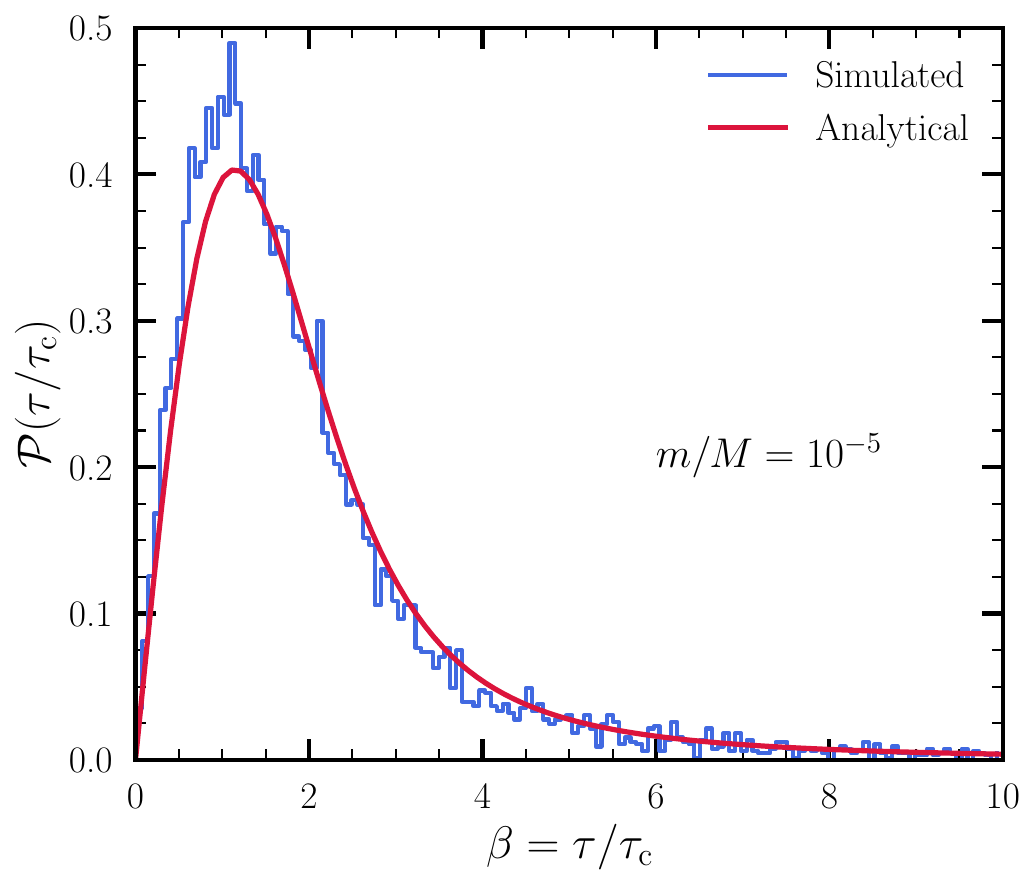}
\includegraphics[width=0.49\textwidth]{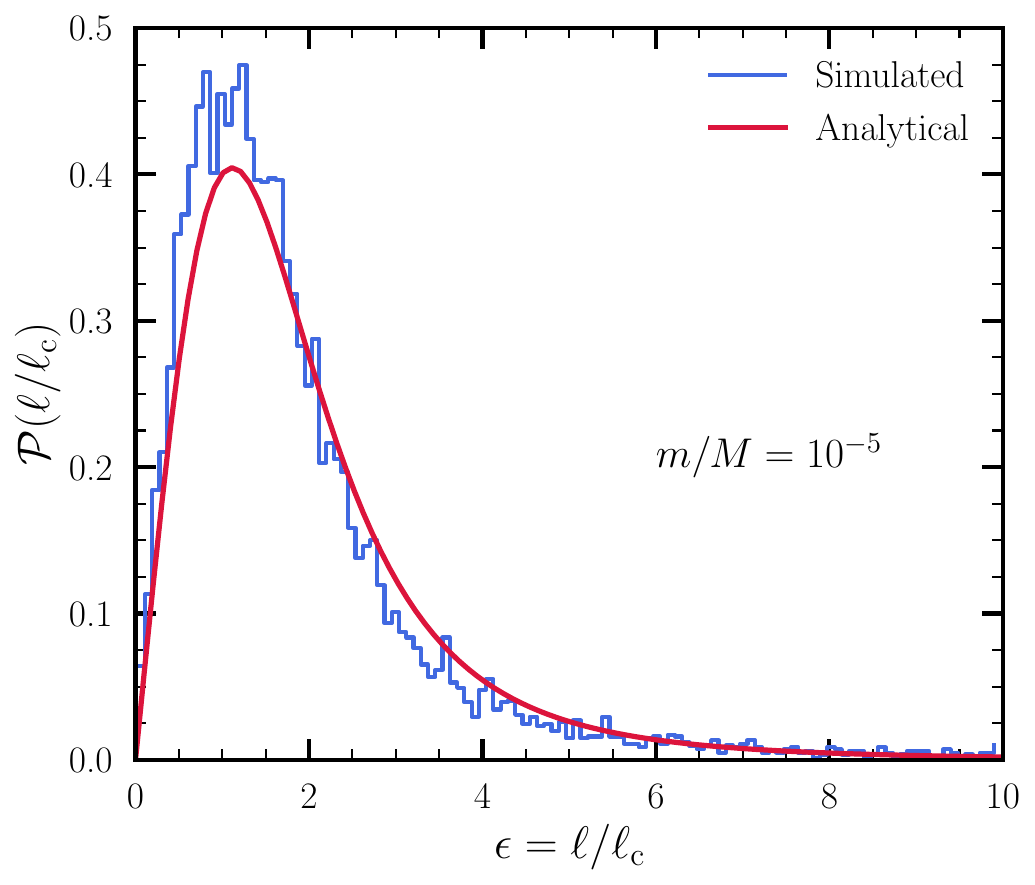}
\caption{The simulated (blue) and analytically estimated (red) probability distribution functions of the torque fraction $\beta$ (\textbf{left panel}) and the normalized angular momentum $\ell/\ell_{\rm c}$ (\textbf{right panel}), evaluated at the distance of closest approach in the simulation.}
\label{fig:beta_l_pdf}
\end{figure}

We now consider the total torque, obtained by summing the contributions from all \lPBHs:
\begin{equation}
\tauvec = -G m^2 \sum_{k} \frac{\rvec \times \rvec_k^\prime}{\left| \rvec - \rvec_k^\prime \right|^3} \, .
\end{equation} 
To determine the probability distribution of the initial torque $\mathcal{P}(\tau_i)$, we integrate over all possible initial configurations of the light PBHs, in analogy to \cref{sec:hPBH_ext} (see Appendix~\ref{app:torque_dist} for details). The resulting probability density function of the torque magnitude $\tau_i$ peaks around the characteristic value
\begin{equation}
\label{eq:tau_c}
    \tau_{\rm c} \simeq 2.6\, G m^2 \, r_{\rm ta} \, \overline{n}_\ell(r_{\rm ta})^{2/3} \,.
\end{equation}
It is then convenient to express the torque in units of this characteristic scale by defining $\beta \equiv \tau_i/\tau_{\rm c}$, whose distribution is
\begin{align}
\label{eq:beta_PDF}
\mathcal{P}(\beta) = 
\beta \int_0^\infty \dd s\, s\, J_0 \left(\beta s\right)
e^{-s^{3/2}} \,,
\end{align}
where $J_0$ denotes the Bessel function of the first kind.
This distribution is illustrated as a red curve in the left panel of Fig.~\ref{fig:beta_l_pdf}.

Once the characteristic scale $\tau_{\rm c}$ is fixed through Eq.~\eqref{eq:tau_c}, the torque distribution for all \lPBHs\ is universally described by $\mathcal{P}(\beta)$. As shown in the left panel of Fig.~\ref{fig:beta_l_pdf}, the distribution peaks around $\beta \sim 1.2$, while values larger than $\beta \sim \mathcal{O}(5)$ are strongly suppressed. Physically, large torques arise only when the individual torque contributions from nearby \lPBHs\ happen to align coherently, thereby reducing the cancellations that typically occur in a random configuration, and are hence less probable. In fact, we find that the total torque is dominated by nearest neighbors: in numerical simulations, we have found that calculating the torque from the 5 nearest \lPBHs\ reduces the median torque by only around 20\%, compared to including all \lPBHs. Given that large value of $\beta$ are suppressed, in the following we adopt $\beta = 1$ as a representative value.

It might be expected that the relative orientation and separation of the \lPBHs~might vary during the infall, hence leading to an evolution of the torque. The evolution of the torque will hence be parametrized through $\alpha$:
\begin{align}
\tau(r) \simeq \tau_i \left(\frac{r_{\rm ta}}{r}\right)^{\alpha}\,.
\end{align}
From Eq.~\eqref{eq:ell_of_tau}, the angular momentum accumulated over the infall due to mutual \lPBH\ interaction is
\begin{align}
\ell = \frac{1}{m \sqrt{2GM}}  \beta \tau_c(r_{\rm ta})\, r_{\rm ta}^{3/2}  \mathcal{B}\left(\frac{3}{2} - \alpha,  \frac{1}{2}\right)\,,
\end{align}
where $\mathcal{B}$ denotes the Euler beta function.

As we argue in Appendix~\ref{app:scale_inv}, on average the torque $\tau(r)$ only varies weakly during the relevant part of the infall, exhibiting at most a mild decrease. This behavior can be understood as the result of two competing effects. On the one hand, the transverse separation of \lPBHs\ tends to be compressed by tidal forces. On the other hand, the radial separation grows during the infall. The net result is a weak radial dependence of the total torque. 
Assuming then that the torque remains roughly constant during the infall ($\alpha \approx 0$), we find
\begin{align}
\label{eq:ell_LL}
\ell =2.89\sqrt{\frac{G}{M}} m \beta  r_{\rm ta}^{5/2{}} \overline{n}_\ell^{2/3} \equiv \beta \ell_{\rm c}\,,
\end{align}
where we have introduced the characteristic angular momentum $\ell_{\rm c}$\footnote{Note that deviations from a constant torque (i.e., $\alpha \neq 0$) primarily affect the normalization of the angular momentum through the $\mathcal{B}$ function, while preserving the scaling of the distribution. Moreover, the numerical value of the $\mathcal{B}$ function varies only by approximately a factor of 2 in the range $\alpha \in [-1,\,1]$.}.

We validate these analytic estimates using $N$-body simulations performed with \texttt{GADGET-4}~\cite{Springel:2020plp}.
The setup is designed to capture the dynamics of the \lPBHs under the combined influence of the central heavy seed and the stochastic torques generated by \lPBH-\lPBH~interactions. Since the simulations cannot resolve the dynamics down to the Schwarzschild radius of the \hPBH, they cannot be used to study direct capture~\cite{Springel:2020plp}. We simulate $N \approx 10^4$--$10^5$ particles evolving in the static gravitational potential generated by a central mass $M$. The light particles are initially distributed according to a density profile $n \propto r^{-9/4}$ and are allowed to infall from rest. We record the angular momentum of the light particles at their distance of closest approach; since a smoothening length must be introduced in the simulation to avoid gravitational divergences, the distance of closest approach represents the last physically valid point of the simulation, after which the simulated particle scatters off and reverses its motion. Simulation details are provided in \cref{app:simulation}. The probability distribution obtained from the numerical simulations is shown as a blue histogram in \cref{fig:beta_l_pdf}.

The right panel of Fig.~\ref{fig:beta_l_pdf} shows the resulting distribution of $\ell/\ell_{\rm c}$ for simulations with mass ratio $m/M = 10^{-5}$. In the analytic treatment, this distribution follows directly from the universal distribution of $\beta$ shown in the left panel, and is found to be in excellent agreement with the simulation results, as in the case of the $\beta$ distribution.
Overall, these results indicate that Eq.~\eqref{eq:ell_LL} provides an accurate description of the final angular momentum, thereby supporting our approximation that the torque remains approximately constant during infall.

Finally, we can derive the critical radius for capture, fixing $\beta = 1$, as:
\begin{align}
\begin{split}
    r_{\rm c}
    & \approx 0.16\,\mathrm{pc}\,f_\ell^{-2/3}\left(\frac{M}{M_\odot}\right)\left(\frac{m}{10^{-16}\,M_\odot}\right)^{-1/3}\\
    &\approx 4.77\,r_\mathrm{sp, eq}\,f_\ell^{-2/3}\left(\frac{M}{M_\odot}\right)^{2/3}\left(\frac{m}{10^{-16}\,M_\odot}\right)^{-1/3}\,.
    \end{split}
\end{align}

\subsection{Torques from heavy PBH motion}
\label{sec:shell_capture}

We now consider the angular momentum generated by the motion of the \hPBH. 
Because of Poisson noise, the \lPBHs distribution within the shells is anisotropic, generating a net force on the \hPBH. If inner shells are captured, the force may change direction, giving rise to a stochastic motion. The resulting displacement of the \hPBH\ from its initial position may eventually become significant, and \lPBHs\ might acquire a net angular momentum.

\begin{figure}[tb]
\centering
\includegraphics[width=0.6\textwidth,trim={1cm} 1cm {1cm} 1cm, clip]{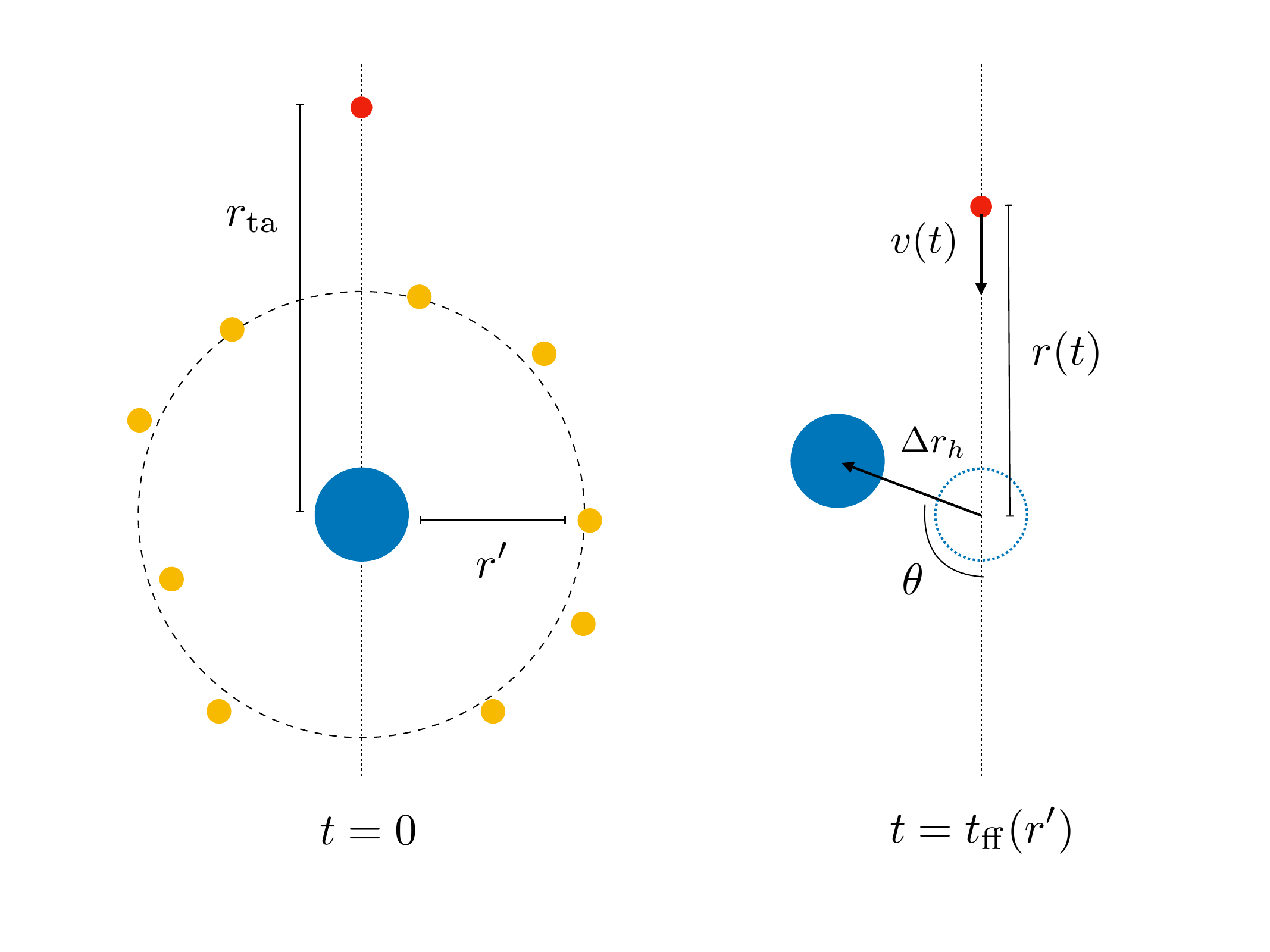}
\caption{We consider that the heavy PBH (blue) captures a shell of light PBHs (yellow) and its position is shifted by $\Delta \mathbf{r}_h$. This means that the velocity $\mathbf{v}$ of the primary light PBH (red) is no longer pointing at the heavy PBH, so the system gains an angular momentum $\mathbf{\ell} = -\Delta \mathbf{r}_h \times \mathbf{v}$.}
\label{fig:setup_hl}
\end{figure}

We calculate the angular momentum $\mathbf{\ell}$ of the \hPBH~with respect to one of the infalling \lPBHs, at an initial radius $\rta$, which we refer to as the \textit{primary} \lPBH~(shown in red in Fig.~\ref{fig:setup_hl}). 

Consider a shell of \lPBHs initially located at radius $r^\prime<\rta$ and accreted after a time $t_\mathrm{ff}(r^\prime)$. Since the force it exerts on the \hPBH during infall grows towards small radii, and the infall velocity \cref{eq:vel_infall} grows rapidly with time, we can assume that most of the displacement of the \hPBH\ happens instantaneously at $t_\mathrm{ff}$.
At this moment, the velocity of the primary \lPBH~is still pointing towards the old position of the \hPBH. This misalignment induces a change in angular momentum of the system $\Delta \mathbf{\ell} = -\Delta \mathbf{r}_h \times \mathbf{v}$, with $\Delta \ell = \Delta r_h v \sin\theta$, as illustrated on the right side of Fig.~\ref{fig:setup_hl}. 

The capture of a single \lPBH\, displaces the position of the \hPBH\ by
\begin{equation}
\Delta \mathbf r_h
\simeq
\frac{m}{M}\,\mathbf r' ,
\end{equation}
which exactly corresponds to the centre-of-mass of the light-heavy binary before capture\footnote{Here, we assume that the displacement of the heavy PBH from the centre of the system is always small compared to the initial \lPBH~positions.}.
Considering now a thin shell of $N$ \lPBHs\ at an initial radius $\mathbf{r}^\prime$, the net displacement after capture will simply be 
\begin{equation}
    \Delta \mathbf{r}_h = \left( \frac{m}{M} \right) \sum_{k = 1}^N \mathbf{r}_k^\prime \, ,
\end{equation}
where we have assumed that the growth in mass of the \hPBH\ due to \lPBH\ capture is negligible. 
As the mean value of $\Delta\ell$ is zero, we calculate the variance of $\Delta \ell$ over realizations:
\begin{equation}
    \Delta \ell^2 = \langle \Delta r_h^2 v^2 \sin^2\theta\rangle = \left(\frac{m}{M}\right)^2 v^2 \left\langle \left|\sum\nolimits_{k = 1}^N \mathbf{r}_k^\prime\right|^2\right\rangle\left\langle \sin^2\theta\right\rangle = \frac{2}{3}N\left(\frac{m}{M}\right)^2 (r^\prime)^2 v^2\,.
\end{equation}
Here, we have made use of the fact that the expectation value $\langle |\sum_{k = 1}^N \hat{\mathbf{n}}_k|^2\rangle = N$, for $N$ randomly oriented unit vectors $\left\{\hat{\mathbf{n}}_k\right\}$.\footnote{
$\langle|\sum_{k=1}^N \hat{\mathbf{n}}_k|^2\rangle 
=  \langle (\sum_{k=1}^N\hat{\mathbf{n}}_k)(\sum_{j=1}^N\hat{\mathbf{n}}_j)\rangle 
= \sum_{k=1}^N\sum_{j=1}^N\left\langle \hat{\mathbf{n}}_k \hat{\mathbf{n}}_j\right\rangle 
=   \sum_{k=1}^N \langle \hat{\mathbf{n}}_k^2\rangle + \sum_{j \neq k } \langle \hat{\mathbf{n}}_k \hat{\mathbf{n}}_j \rangle = N\,.$}
We also note that the direction of $\Delta \mathbf{r}_h$ is randomly oriented in 3-dimensions, meaning that the angle $\theta \in [0, \pi]$ is distributed as a polar angle, with $\langle \sin^2\theta\rangle = 2/3$.

Assuming that the displacements induced by different shells are uncorrelated, the corresponding variances add, giving
\begin{equation}
    \Delta \ell^2_\mathrm{tot} = \frac{8\pi}{3}\frac{m}{M^2}\int_{\rS}^{r_{\rm ta}} (r^\prime)^4 \rho(r^\prime) v_r^2(t)\,\mathrm{d}r^\prime\, ,
    \label{eq:ell_HL}
\end{equation}
where $v_r$ is the infall velocity given by \cref{eq:vel_infall}.\footnote{Here, we integrate from the Schwarzschild radius upwards, but in practice we can safely take the lower limit of this integral to zero. The contribution from shells captured at small initial radii is strongly suppressed by the fact that the radial velocity is initially zero and grows slowly with time.} The density is assumed not to diverge significantly from the initial profile from \cref{eq:rho_ta}, so we consider  $n(r') \propto (r')^{-9/4}$. The characteristic angular momentum will hence be
\begin{equation}
\label{eq:delta_ell_h}
     \ell_{\rm c} = \sqrt{\Delta \ell_\mathrm{tot}^2} \approx 4.5 \sqrt{\frac{G}{M}}m \,\left(\overline{n}_\ell(r_{\rm ta})\right)^{1/2}\, r_{\rm ta}^2\,.
\end{equation}

We validate this calculation by comparing with numerical simulations. We use the code \texttt{NbodyIMRI}~\cite{NbodyIMRI}, which uses a leapfrog algorithm to follow the Newtonian motion of light particles around a central black hole, neglecting the pairwise interactions between light particles (see Ref.~\cite{Kavanagh:2024lgq} for full details). We simulate a central particle of mass $M = 1 \,M_\odot$, surrounded by $N = 10^3$ particles of mass $m$, initially at rest and distributed following a $n \propto r^{-9/4}$ density profile. The light particles are captured when they cross the Schwarzschild radius of the \hPBH, and we record the final angular momentum of each \lPBH~at the moment of capture. 

\begin{figure}[tb!]
\centering
\includegraphics[width=0.48\textwidth]{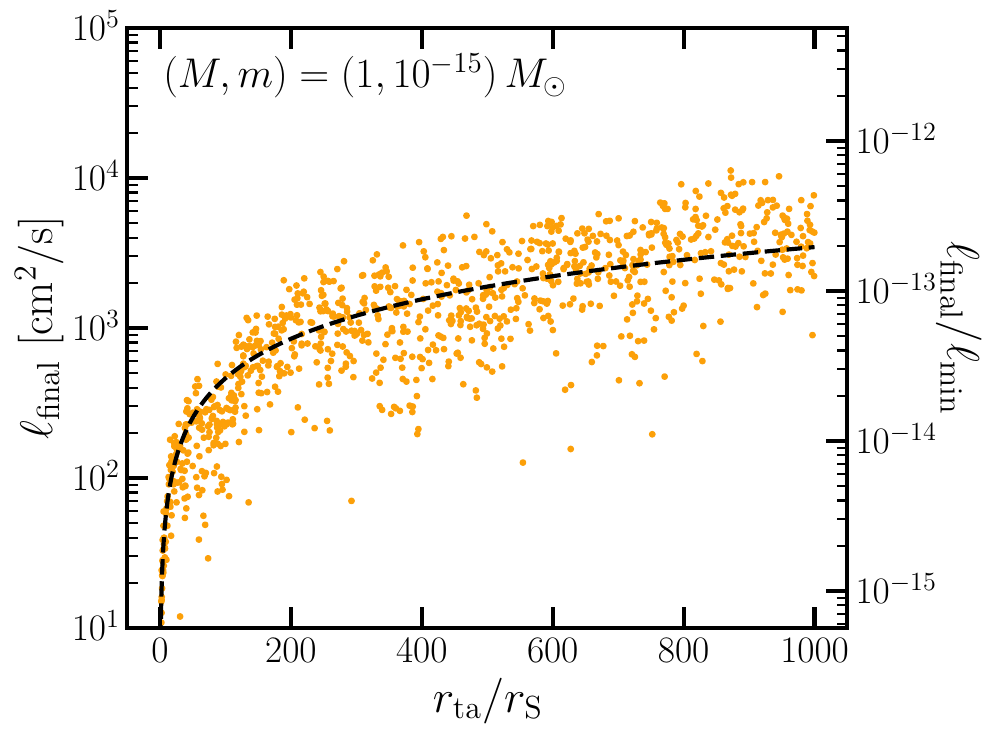}
\includegraphics[width=0.48\textwidth]{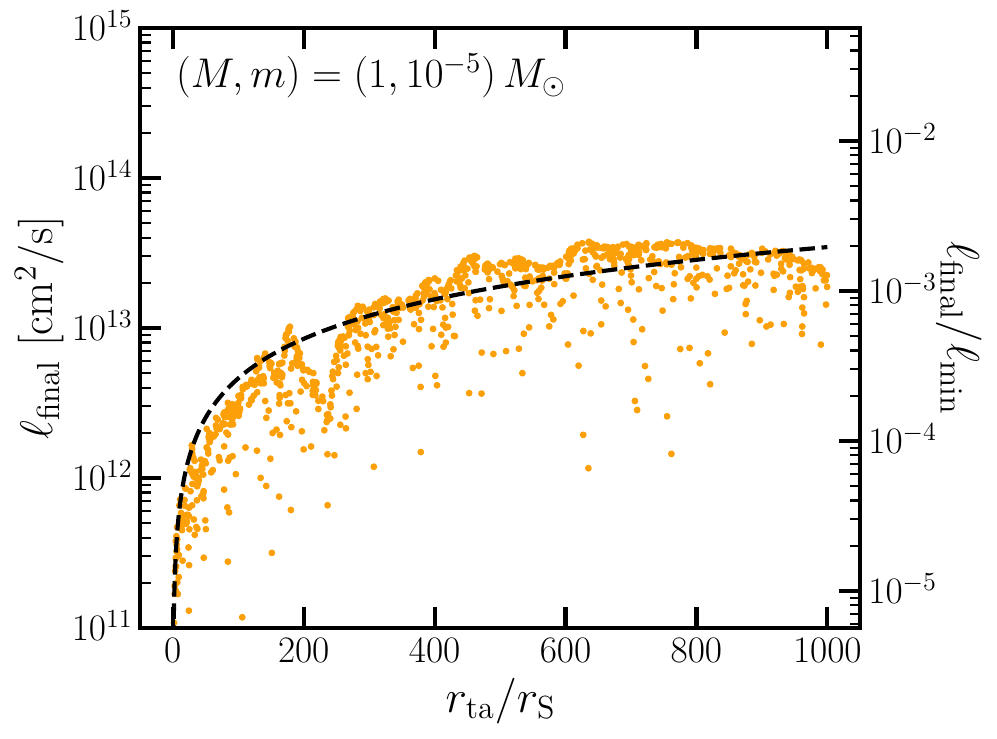}
\caption{Simulation results using \texttt{NbodyIMRI}~\cite{Kavanagh:2024lgq,NbodyIMRI} for the final angular momentum of particles of mass $m$ infalling towards a heavy central object of mass $M \approx M_\odot$. The particles are initially distributed with a density profile $n \propto r^{-9/4}$ in a sphere of radius $r_\mathrm{max} = 10^3\,\rS$, where $\rS$ is the Schwarzschild radius of the heavy black hole. The light particles are captured when they pass within $r < \rS$ of the heavy PBH (which is dynamical in the simulations). The dashed black line corresponds to the estimate of Eq.~\eqref{eq:delta_ell_h}. The left panel considers the particles to have a mass of $m = 10^{-15} M_\odot$, the right panel $m = 10^{-5} M_\odot$.}
\label{fig:simulations_HL}
\end{figure}

The results are shown in Fig.~\ref{fig:simulations_HL} for mass ratios $m/M = 10^{-15}$ (left) and $10^{-5}$ (right). In both cases, the analytic estimate (dashed black line) provides a good prediction of the final angular momentum over the scales probed by the simulations. The spread in the points demonstrates that the full distribution of angular momentum varies by no more than $\approx O(1)$ of magnitude around the characteristic  value $\ell_{\rm c}$ we estimated. For larger mass ratios (right panel), we see that the black dashed line presents a mild over-estimate of the final angular momentum, as the \lPBH-induced motion of the \hPBH starts to become significant, and our assumption that the displacement of the \hPBH\ is small begins to break down.

Through the characteristic angular momentum of Eq.~\eqref{eq:delta_ell_h}, we can derive the critical radius for capture:
\begin{align}
\begin{split}
    r_{\rm c}
    &\approx 1.43\times 10^2 \,\mathrm{pc} \times f_\ell^{-4/7}\left(\frac{M}{1\,M_\odot}\right)^{9/7}\left(\frac{m}{10^{-16}\,M_\odot} \right)^{-4/7}\\
    &\approx 4.2 \times 10^3 \,\,r_{\mathrm{sp, eq}}  \times f_\ell^{-4/7}\left(\frac{M}{1\,M_\odot}\right)^{20/21}\left(\frac{m}{10^{-16}\,M_\odot} \right)^{-4/7} \,.
    \end{split}
\end{align}

\section{Results}
\label{sec:results}
\begin{figure}[tb]
\centering
\includegraphics[width=\textwidth]{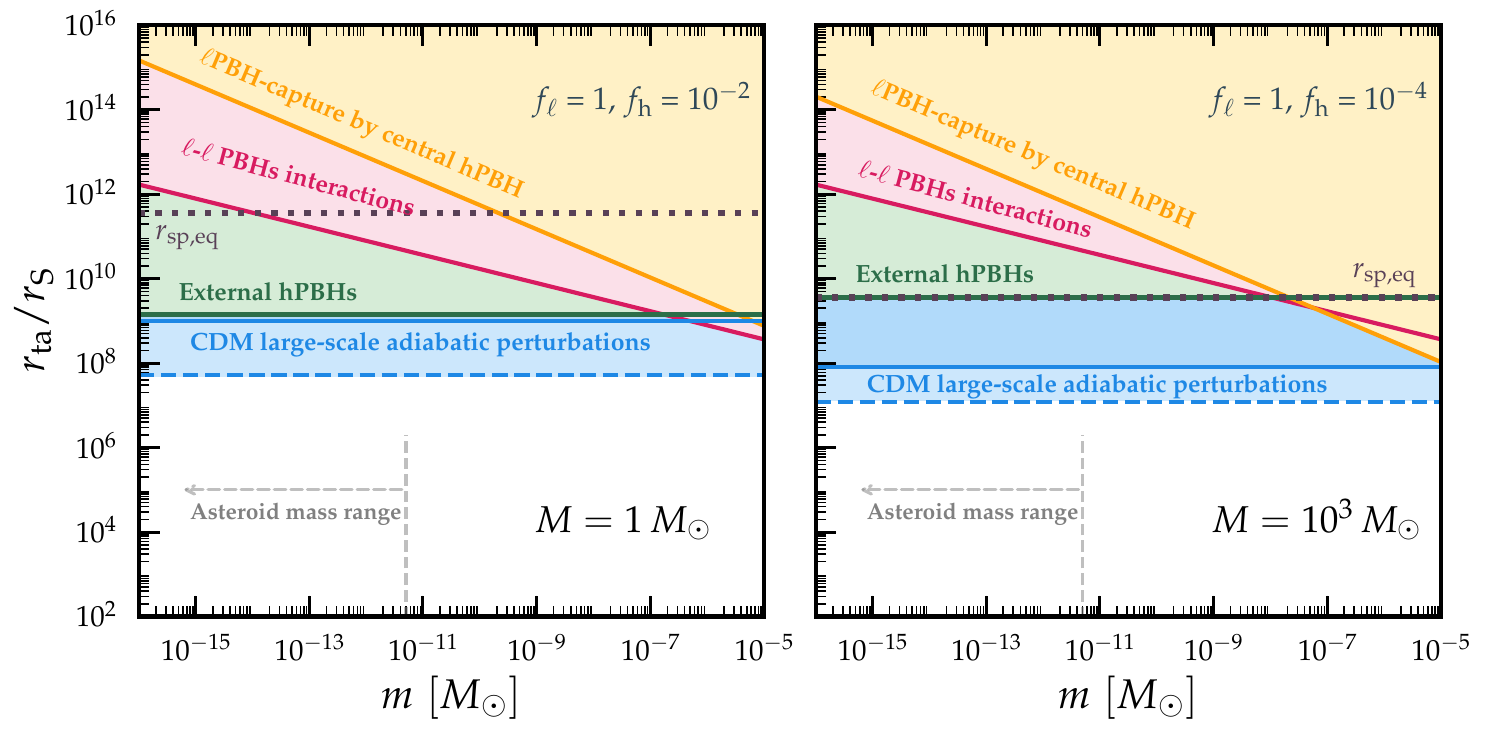}
\caption{Critical turn-around radius (normalized to the Schwarzschild radius $r_\mathrm{S}$ of the \hPBH) as a function of the \lPBH~mass $m$, for each 
torque mechanism discussed in the text: external \hPBHs~(\textcolor{plotgreen}{green}, \cref{sec:hPBH_ext}); large-scale 
adiabatic CDM perturbations (\textcolor{plotblue}{blue}, 
\cref{sec:adiabatic}; solid: Planck-extrapolated PPS, dashed: enhanced 
PPS); \lPBH--\lPBH~interactions (\textcolor{plotred}{magenta}, 
\cref{sec:ell-ell}); and capture of \lPBH~shells by the \hPBH~(\textcolor{plotorange}{orange}, \cref{sec:shell_capture}). 
For each mechanism, the shaded region above the corresponding line 
indicates where the induced angular momentum is sufficient for 
infalling \lPBHs~to avoid capture. Shells turning around within the white area are \textit{swallowed} by the \hPBH. Left and right panels correspond 
to $M = M_\odot$ and $M = 10^3\,M_\odot$, respectively, with 
$f_\ell = 1$, $f_h = 10^{-2}$ (left) and $f_h = 10^{-4}$ (right). The gray dashed lines indicate the PBH mass window that is still viable to constitute the totality of DM. }
\label{fig:bounds}
\end{figure}

Finally, we are able to map out the regions of parameter space 
where each torque mechanism generates sufficient angular momentum for infalling \lPBHs~to avoid capture. These are depicted in \cref{fig:bounds} as a function of their turn-around radius $r_\mathrm{ta}$ (normalized to the \hPBH~Schwarzschild radius $\rS$) and their mass $m$. The left and right panels show results for \hPBH masses of $1 \, \Msun$ and $10^3 \, \Msun$, respectively. For both $M$ benchmarks we fix $f_\ell = 1$, with $f_h = 10^{-2}$ for $M = 1 \,\Msun$ and $f_h = 10^{-4}$ for $M = 10^3 \Msun$, according to current constraints~\cite{Mroz:2024wag, Andres-Carcasona:2026avd, Agius:2024ecw}.
%~\cite{Green:2020jor,PBHbounds}. 
Each of the thick solid lines and colored regions corresponds to one of the mechanisms described in the previous sections: torques from external \hPBHs\ (\textcolor{plotgreen}{green}, Sec.~\ref{sec:hPBH_ext}); torques due to large-scale CDM perturbations (\textcolor{plotblue}{blue}, Sec.~\ref{sec:adiabatic}); torques from \lPBH-\lPBH~interactions within the spike (\textcolor{plotred}{magenta}, Sec.~\ref{sec:ell-ell}); and torques due to the capture of \lPBHs by the \hPBH~(\textcolor{plotorange}{orange}, Sec.~\ref{sec:shell_capture}).
The shaded regions correspond to $r_\mathrm{ta} > r_{\rm c}$, where the corresponding mechanism is expected to generate enough angular momentum for the infalling \lPBHs\ to avoid capture. 
To provide a reference scale for the size of the spike, the turn-around radius at equality, $r_{\mathrm{sp},\mathrm{eq}}$ (\cref{eq:r_sp_eq}), is shown as a horizontal gray dotted line. 

Across both benchmark values of \hPBH~mass $M$, and the full range of \lPBH~masses $m$, we find that the largest angular momenta are induced by tidal torques due to the large-scale adiabatic component of CDM perturbations. Tidal torques from the other \hPBHs~are subdominant to this contribution for the DM fractions allowed by current constraints.
Torques generated from small-scale anisotropies in the \lPBH\ distribution become increasingly important as the \lPBH\ mass increases, as expected. For $M = 1 \, \Msun$, \lPBH-\lPBH\ interactions become sufficiently strong above $m \sim 10^{-12 } \Msun$ to allow a fraction of the \lPBHs enclosed within $r_{\mathrm{sp},\mathrm{eq}}$ to escape capture. For a heavier \hPBH~with $M = 10^3 \, \Msun$, these only provide sufficient torques above $m \sim 10^{-8} \Msun$, where they are subdominant compared to the torques induced by \lPBH\ capture from the central \hPBH. 

In the white regions instead, where $r_\mathrm{ta} < r_{\rm c}$, none of the mechanisms studied here generates sufficient angular momentum to avoid capture. In other words, the white region represents the part of the spike that would be \textit{swallowed} by the \hPBH.  
Estimating the torques conservatively from the Planck-extrapolated PPS, we find that the captured shells are those reaching turn-around at radii smaller than $r_\mathrm{c} \sim 10^{8} - 10^9\,\rS$. When considering the enhanced PPS, the critical radius is reduced by about an order of magnitude.
The captured \lPBHs constitute at most a few percent of the total \lPBH~population enclosed within the turn-around radius at equality, dropping to $\sim 0.1\%$ for $M = 1\,M_\odot$ and $\sim 1\%$ for $M = 10^3\,M_\odot$ when using the enhanced primordial spectrum.
These innermost shells, however, would represent the densest and most phenomenologically relevant part of the spike~\cite{Coogan:2021uqv,Feng:2024obn}.

We expect the \lPBHs which avoid capture to settle into elliptical orbits around the central \hPBH, building up the spike -- similarly to the WIMP spike formation scenario described in \cref{sec:intro}~\cite{Eroshenko:2016yve,Carr:2020mqm}.
In particular, those that only marginally escape capture should settle into highly eccentric orbits, contributing to the density at small radii. Nevertheless, we expect the resulting inner core to be significantly less dense than those obtained in WIMP scenarios.
We leave a detailed calculation of the resulting density profile, and the associated phenomenology, for future work.

\begin{figure}
    \centering
    \includegraphics[width=\linewidth]{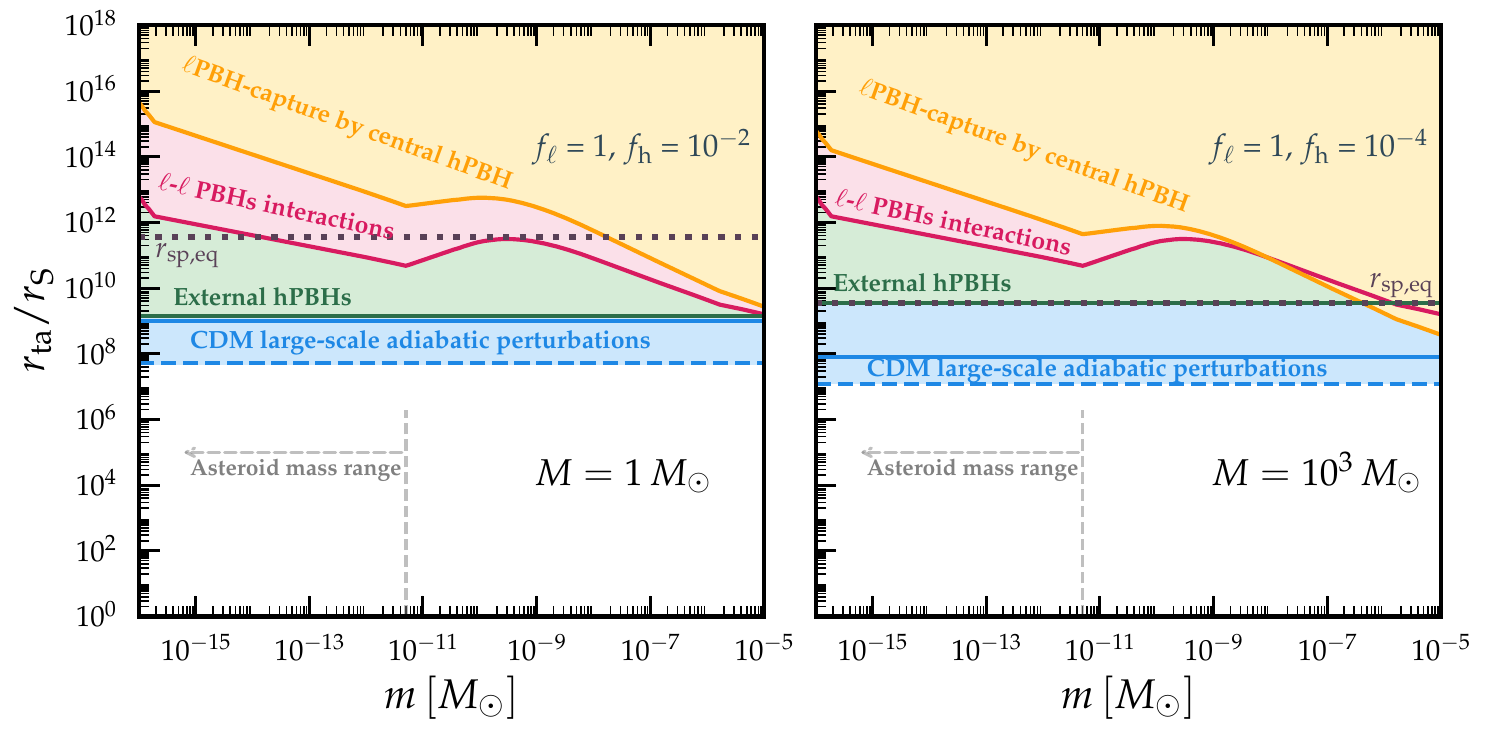}
    \caption{Same as Fig.~\ref{fig:bounds}, but fixing $f_\ell$ to the maximum allowed by constraints from evaporation and microlensing~\cite{PBHbounds}
    }
    \label{fig:bounds_fpbh}
\end{figure}

So far, we have assumed that \lPBHs contribute almost all of the DM density. However, this is only possible for PBHs in the asteroid mass range $m \in [10^{-16},10^{-11}]\,M_\odot$~\cite{Green:2020jor,Tinyakov:2024mcy}, which is marked in the figures by a vertical gray dashed line. Below this range, the Hawking radiation from light PBHs leads to observable signatures, while above this mass range, the PBHs are heavy enough to be observed in microlensing surveys. In Fig.~\ref{fig:bounds_fpbh}, we show the regions where \lPBHs can avoid capture, fixing $f_\ell$ to the maximum allowed by these constraints, while keeping $f_\ell = 1$ inside the asteroid mass window. Above the asteroid mass window, $f_\ell$ is constrained to be less than $10^{-3}$--$10^{-2}$, substantially reducing the allowed density of \lPBHs and thus reducing the angular momentum produced by $\ell$--$\ell$ interactions and \lPBH~capture by the \hPBH. 
Also in this case, large-scale CDM perturbations remain the most significant mechanism to produce angular momentum.

In this work, we have treated each torque-generating mechanism independently. Near the crossover regions in Figs.~\ref{fig:bounds} and \ref{fig:bounds_fpbh}, however, two mechanisms can contribute comparably to the generation of angular momentum. Their combined effect may therefore be larger than that of either mechanism acting alone, potentially allowing shells at smaller radii to avoid capture. This consideration becomes even more relevant in the upper regions of these plots, where several mechanisms may contribute simultaneously and their interplay may need to be treated self-consistently. 
Given that the contributions from internal and external torques are expected to be uncorrelated, cancellations at large radii from the superposition of different mechanisms are very unlikely. 

In addition, we have assumed \lPBHs~to be Poisson distributed at formation. In realistic formation scenarios, however, PBH positions are expected to be correlated (see, e.g.~\cite{Desjacques:2018wuu,Suyama:2019cst,Young:2019gfc}). Such initial clustering would enhance both \lPBH--\lPBH~torques and tidal torques from neighboring \hPBHs, potentially increasing the relative importance of these mechanisms and further reducing the size of the captured region. These assumptions therefore render our results as conservative upper bounds on the size of the captured region.

\section{Conclusions}
\label{sec:conclusions}

In this work, we have considered a scenario in which light primordial black holes (\lPBHs) constitute the bulk of the dark matter, while a subdominant population of heavier primordial black holes (\hPBHs) acts as seeds for the formation of overdense structures, or \textit{spikes}. Assuming  that PBHs are formed with no peculiar velocities, a necessary condition for spike formation is that infalling \lPBHs~acquire sufficient angular momentum to avoid direct capture by the central \hPBH.

We identified and quantified several mechanisms through which 
\lPBHs~of mass $m$ can acquire this angular momentum around a central \hPBH~of mass $M$: tidal torques from large-scale adiabatic CDM perturbations, tidal torques from neighboring \hPBHs, mutual \lPBH--\lPBH~interactions, and the stochastic recoil of the \hPBH~due to \lPBH~capture. For each mechanism, we derived a critical turnaround radius $r_{\rm c}$, below which infalling \lPBHs~are expected to be captured by the central \hPBH. Across the mass ranges considered, the dominant source of angular momentum is the tidal torque from large-scale adiabatic CDM perturbations. Since this torque is independent of the 
\lPBH~mass $m$ at fixed turnaround radius, the size of the captured region is set primarily by the \hPBH~mass $M$ and by the primordial power spectrum of CDM perturbations: a Planck-extrapolated spectrum results in a captured region extending up to $r_{\rm c} \sim 10^{9}\,\rS$, while an enhanced spectrum, required to accommodate PBH formation, reduces this to $r_{\rm c} \sim 10^{7}\,\rS$, where $\rS$ is the Schwarzschild radius of the \hPBH. Other mechanisms, in particular \lPBH--\lPBH~interactions and \hPBH~recoil, become relevant at larger \lPBH~masses even though their contribution to the angular momentum remains subdominant in almost all cases.

In all scenarios studied, the captured shells constitute at most a few percent of the total \lPBH~mass enclosed within the spike radius at equality. However, since these shells correspond to the highest-density regions of the spike, their removal is expected to significantly affect the resulting density profile, leaving the inner region less dense than the canonical $\rho \propto r^{-9/4}$ profile found for particle DM. We note that many interesting probes of DM spikes, such as their environmental effect imprinted in gravitational wave signals~\cite{Eda:2013gg,Eda:2014kra,Bertone:2019irm,Cardoso:2019rou,Kavanagh:2020cfn,Feng:2024obn}, are only observable if the inner spike reaches very large densities $\gtrsim \mathcal{O}(10^{20})\,M_\odot\,\mathrm{pc}^{-3}$~\cite{Coogan:2021uqv,Cole:2022ucw}.  While we leave a detailed calculation of the resulting density profile to future work, it is clear that any suppression of the inner density -- such as that suggested here -- could have drastic effects on the detectability of \lPBH\ spikes.

This work represents a first step toward understanding the formation of \lPBH~spikes in the early Universe, and several aspects of their subsequent evolution remain to be explored. \lPBHs~that only marginally avoid capture settle into highly eccentric orbits, which may generate a distinctive gravitational-wave signal; this could drive their inspiral and eventual merger with the \hPBH~on timescales much shorter than the age of the Universe. Mutual \lPBH--\lPBH~interactions within the spike may further drive its evolution through additional gravitational-wave emission and \lPBH-\lPBH~coalescence, potentially triggering a hierarchical merger process. For the heavier \hPBH~benchmark, this additional growth channel could help provide the rapid mass growth required for $M \sim 10^3\,M_\odot$ seeds to act as viable progenitors of supermassive black holes. We leave a detailed study of these dynamical processes, and their observational consequences, to future work.

\section*{Acknowledgments}
We are grateful to Dominic Agius for useful discussions and collaboration in the early stages of this work. We also thank Aurelio Amerio, Gianfranco Bertone and Sefa Pamuk for helpful discussions. AT is also grateful to the members of the astroparticle and cosmology group at LAPTh for their kind hospitality and insightful conversations during the final stages of this work.

VDR and AT acknowledge financial support by the grant CIDEXG/2022/20 (from Generalitat Valenciana) and by the Spanish grants CNS2023-144124 (MCIN/AEI/10.13039/501100011033 and “Next Generation EU”/PRTR), PID2023-147306NB-I00, and CEX2023-001292-S (MCIU/AEI/ 10.13039/501100011033).

BJK and FS acknowledge funding from the \textit{Consolidaci\'on Investigadora} Project \textsc{DarkSpikesGW}, reference CNS2023-144071, financed by MCIN/AEI/10.13039/501100011033 and by the European Union ``NextGenerationEU"/PRTR.

BJK also acknowledges support from the project SA101P24 (Junta de Castilla y León).

DG acknowledges support from the Research grants TAsP (Theoretical Astroparticle Physics) and TEONGRAV funded by INFN.

\appendix

\section*{Appendices}
\addcontentsline{toc}{section}{Appendices}

\section{Torque distribution from random flights}
\label{app:torque_dist}

The Markov theory of random flights~\cite{DelPopolo:1998in, Chandrasekhar:1943ws, Ali-Haimoud:2017rtz} provides a framework for computing the probability distribution of a vector quantity that arises as the sum of many independent, identically distributed contributions. In this appendix, we apply this formalism to derive the 
torque distributions in \cref{sec:hPBH_ext,sec:ell-ell}.

\subsection*{The Markov method}
\label{app:subsec_Markov}

Following~\cite{Chandrasekhar:1943ws}, consider a set of $N$ independent $n$-dimensional vectors ${\mathbf{\phi}_j}$ and define their sum as $\mathbf{\Phi} = \sum_{j=1}^N \mathbf{\phi}_j$. Each vector $\mathbf{\phi}_j$ depends on a random three-dimensional position vector $\mathbf{r}_j$. The $\mathbf{r}_j$'s are independent and identically distributed, drawn from some normalized probability density $f(\mathbf{r})$.  The probability density for $\mathbf{\Phi}$ to take the value $\mathbf{\Phi}_0$ is
\begin{equation}
\mathcal{P}(\mathbf{\Phi}_0)
= \int \prod_{j=1}^N d^3 \mathbf{r}_j \;
\delta^{(n)}\!\big(\mathbf{\Phi}( \mathbf{r}_j) - \mathbf{\Phi}_0\big)
%\prod_{j=1}^N
f(\mathbf{r}_j)\,.
\end{equation}
Using the Fourier representation of the $n$-dimensional Dirac delta, we obtain
\begin{equation}
\mathcal{P}(\mathbf{\Phi}_0)
=
\frac{1}{(2\pi)^n}
\int d^n\mathbf{k}\;
e^{-i\mathbf{k}\cdot\mathbf{\Phi}_0}
\prod_{j=1}^N
\int d^3\mathbf{r}_j\,
f(\mathbf{r}_j)\,
e^{i\mathbf{k}\cdot\mathbf{\phi}_j(\mathbf{r}_j)}
\equiv
\frac{1}{(2\pi)^n}
\int d^n\mathbf{k}\;
e^{-i\mathbf{k}\cdot\mathbf{\Phi}_0}
\mathcal{C}_N(\mathbf{k})\, .
\end{equation}
The characteristic function reduces to
$\mathcal{C}_N(\mathbf{k}) = [\mathcal{I}(\mathbf{k})]^N$,
where
$\mathcal{I}(\mathbf{k}) = \int d^3\mathbf{r}\,f(\mathbf{r})\,
e^{i\mathbf{k}\cdot\mathbf{\phi}(\mathbf{r})}$.
Writing $\mathcal{I} = 1 - \mathcal{J}/N$ with
\begin{equation}
\mathcal{J}(\mathbf{k})
=
N \int d^3\mathbf{r}\,
f(\mathbf{r})
\left(1-e^{i\mathbf{k}\cdot\mathbf{\phi}(\mathbf{r})}\right),
\end{equation}
and taking $N\to\infty$ at fixed $\mathcal{J}$, one obtains
$\mathcal{C}_N =(1- \mathcal{J}/N)^N\to e^{-\mathcal{J}(\mathbf{k})}$, so that
\begin{equation}
\mathcal{P}(\mathbf{\Phi}_0)
=
\frac{1}{(2\pi)^n}
\int d^n\mathbf{k}\;
\exp\!\left[-i\,\mathbf{k}\cdot \mathbf{\Phi}_0 - \mathcal{J}(\mathbf{k})\right].
\end{equation}
This is the central result of the Markov method of random flights. 

\subsection*{Application to the torque from neighbor \hPBHs}

Let us now consider the tidal torques produced by neighboring \hPBHs~on the light-heavy binary, discussed in~\cref{sec:hPBH_ext}. Following \cite{Ali-Haimoud:2017rtz}, the torque per unit \lPBH~mass exerted by a single external \hPBH~is
\begin{align}
   \mathbf{\tau}' \equiv \frac{\mathbf{\tau}}{m} 
    =
    \frac{3 G M}{R^3}\,r^2
    \left(\hat{\mathbf{r}}\times \hat{\mathbf{R}}\right)
    \left(\hat{\mathbf{R}}\cdot \hat{\mathbf{r}}\right) \, ,
\end{align}
where $R = |\mathbf{R}|$ is the position of the external \hPBH~relative to the central one, and $r = |\mathbf{r}|$ is the \lPBH-\hPBH~separation (see \cref{fig:hPBH_torques}).

To evaluate $\mathcal{J}$, we introduce a rotated frame defining 
new angular coordinates $(\theta', \phi)$ via $\hat{\mathbf{R}} = \left(\hat{\mathbf{R}}' \cdot \hat{\mathbf{r}}\right)\hat{\mathbf{r}} + \hat{\mathbf{r}}\times\hat{\mathbf{R}}'$, where $\hat{\mathbf{R}}'$ denotes the unit vector in the rotated frame, with 
$\theta'$ the polar angle from $\hat{\mathbf{\sigma}} = \hat{\rvec}_\ell\times \hat{\mathbf k}$ and $\phi$ the azimuthal angle measured from $\hat{\mathbf k}$ in the $(\hat{\mathbf{k}}, \hat{\mathbf{r}})$ plane. Note that 
$|\mathbf{R}'| = |\mathbf{R}| = R$ since this is a pure rotation. Moreover, the new angle $\theta'$ is related to $\theta$ used in \cref{sec:hPBH_ext}  through $\langle\sin^2\theta'\cos 2\phi - \sin^4\theta'\cos^4\phi\rangle = \frac{1}{4}\langle\sin^2\theta\rangle$.

\begin{figure}[tb]
\centering
\begin{tikzpicture}[scale=2, >=stealth]
% Axes
\draw[->, thick] (0,0) -- (1.4,0) node[right] {$\hat{\mathbf k}$};
\draw[->, thick] (0,0) -- (0,1.6) node[above] {$\hat{\mathbf\sigma} = \hat{\rvec}_\ell\times \hat{\mathbf k}$};
\draw[->, thick] (0,0) -- (-0.9,-0.8) node[below left] {$\hat{\mathbf r}$};

% Vector R'
\coordinate (r) at (0.8,1.1);
\draw[->, thick, red] (0,0) -- (r) node[above right] {$\mathbf R'$};

% Projections
\draw[dashed, red] (r) -- (0.8,-0.25);
\draw[dashed, red] (0,0) -- (0.8,-0.25);

% Angles
\draw (0,0.6) arc[start angle=90,end angle=53,radius=0.6];
\node at (0.35,0.75) {$\theta'$};

\draw (0.4,0) arc[start angle=0,end angle=-18,radius=0.4];
\node at (0.3,-0.18) {$\phi$};
\end{tikzpicture}
\caption{Schematic representation of the coordinate system used in the external torque calculation.}
\label{fig:other_axis_torque}
\end{figure}
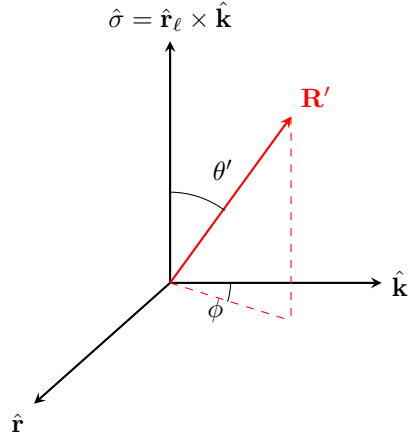

For a homogeneous Poisson distribution of \hPBHs,  with number density 
$n_h$ and $f(\mathbf{R}) = 1/V$, the Markov integral gives
\begin{equation}
    \mathcal{J}
    =
    n_h \int d^3 \mathbf{R}'
    \left[
    1
    -
    \exp\left(
    - i\, r^2\, k\,
    \frac{3 G M}{2R^3}
    \sin 2\phi\,\sin^2\theta'
    \right)
    \right]
    =
    \frac{4\pi}{3}\,
    n_h\,G\,M\,r^2\,k.
\end{equation}

Unlike the $\ell$--$\ell$ interaction case, which we will discuss shortly, where $\mathcal{J} \propto k^{3/2}$, here  $\mathcal{J}$ is exactly linear in $k$. This reflects the slower fall-off of the tidal field ($\propto R^{-3}$) compared to the direct force ($\propto r^{-2}$).
The probability distribution per unit of \lPBH~mass for the magnitude of the torque induced by $N$ external \hPBHs\ is therefore
\begin{align}
\label{eq:pof_h_torques_integ}
\mathcal P\!\left(\tau_0'\right)
&=
\frac{\tau_0'}{(\tau_c^h)^2}
\int_0^\infty ds\, s\,
J_0\!\left(\frac{\tau_0'}{\tau_c^h}s\right)
e^{-s} \,,
\end{align}
with characteristic torque per unit mass
\begin{align}
\dfrac{\tau_c^h}{m}
=
\frac{4\pi}{3}\,
n_h\,G\,M\,r^2
=\frac{G M  r^2}{\bar{R}^3} \,,
\end{align}
%\frac{4\pi}{3}\,f_h\,\rho_{\rm eq}^{\rm DM}
%\left(\frac{a_{\rm eq}}{a}\right)^3
%G\,r^2 \, ,
%\end{align}
where in the second equality we have used $\bar R=(3/4 \pi n_h)^{1/3}$. Notice that this characteristic value corresponds to the angle-averaged torque from the nearest neighbor, derived in \cref{sec:hPBH_ext}. The integral in \cref{eq:pof_h_torques_integ} can be computed analytically and gives
\begin{equation}
    \mathcal{P}(\beta) = \dfrac{\beta}{\left( 1+\beta^2\right)^{3/2}} \,,
\end{equation}
with $\beta \equiv \tau_0/\tau_\mathrm{c}$. 

\subsection*{Application to the torque from \lPBH-\lPBH~interactions}
\label{app:subsec_ll}

Consider the torque, computed with respect to the central \hPBH, exerted on a \lPBH~at position $\mathbf r_\ell$ by a distribution of \lPBHs at positions $\mathbf r_p$
\begin{align}
\mathbf{\tau}_0
= \sum_p \mathbf{\tau}_p
= \mathbf{r}_\ell \times \sum_p \mathbf{F}_p
= Gm^2 \sum_p
\frac{\mathbf{r}_\ell \times (\mathbf{r}_\ell - \mathbf{r}_p)}
{|\mathbf{r}_\ell - \mathbf{r}_p|^3}
= -Gm^2 \sum_p
\frac{\mathbf{r}_\ell \times \mathbf{r}_p}
{|\mathbf{r}_\ell - \mathbf{r}_p|^3},
\end{align}
where $\mathbf{F}_p = Gm^2(\mathbf{r}_\ell - \mathbf{r}_p)/
|\mathbf{r}_\ell - \mathbf{r}_p|^3$ is the gravitational force on the 
\lPBH~due to perturber $p$, and we used $\mathbf{r}_\ell \times 
\mathbf{r}_\ell = 0$ in the last passage.
Since $\mathbf{\tau}_p \propto \mathbf{r}_\ell \times \mathbf{r}_p$, each individual torque is perpendicular to $\mathbf{r}_\ell$, and hence confined to the two-dimensional plane orthogonal to $\hat{\mathbf{r}}_\ell$. The total torque $\mathbf{\tau}_0$ is therefore a two-dimensional vector in this plane, and the Fourier variable $\mathbf{k}$ in the Markov integral is correspondingly two-dimensional. 

\begin{figure}[t!]
\centering
\begin{tikzpicture}[scale=2, >=stealth]
% Axes
\draw[->, thick] (0,0) -- (1.4,0) node[right] {$\hat{\mathbf r}_\ell$};
\draw[->, thick] (0,0) -- (0,1.6) node[above] {$\hat{\mathbf k}$};
\draw[->, thick] (0,0) -- (-0.9,-0.8) node[below left] {$\hat{\mathbf\sigma} = \hat{\rvec}_\ell\times \hat{\mathbf k}$};

% Vector r_\ell
\coordinate (r) at (0.8,1.1);
\draw[->, thick, red] (0,0) -- (r) node[above right] {$\mathbf x{'}$};

% Projections
\draw[dashed, red] (r) -- (0.8,-0.25);
\draw[dashed, red] (0,0) -- (0.8,-0.25);

% Angles
\draw (0,0.6) arc[start angle=90,end angle=53,radius=0.6];
\node at (0.35,0.75) {$\theta$};

\draw (0.4,0) arc[start angle=0,end angle=-18,radius=0.4];
\node at (0.3,-0.18) {$\phi$};
\end{tikzpicture}
\caption{Schematic representation of the coordinate system used in the $\ell$--$\ell$ torque calculation of \cref{app:subsec_ll}.}
\label{fig:axis_torque}
\end{figure}
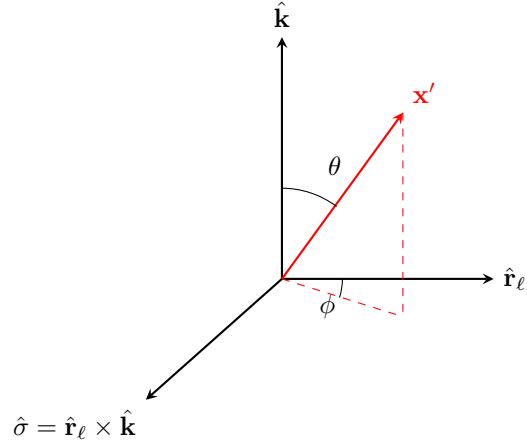

Next, we perform the substitution $\mathbf{r} = \mathbf{r}_\ell - \mathbf{x}$ 
and then decompose $\mathbf{x} = (\mathbf{x}'\cdot\hat{\mathbf r}_\ell)\,\hat{\mathbf r}_\ell +\hat{\mathbf r}_\ell\times\mathbf{x}'$, introducing spherical coordinates $(x', \theta, \phi)$ (see Fig.~\ref{fig:axis_torque}).

In the limit $x' \ll r_\ell$, the single-particle torque magnitude reduces to
$|\mathbf{\tau}| \approx Gm^2 r_\ell \sin\theta / x'^2$,
and $\mathcal{J}$ becomes
\begin{align}
\mathcal{J} =
N \int d^3\xvec{'} \; f(\rvec_\ell - \mathbf{x}{'})
\left( 1 - \exp\left\{-i\,k\, G\,m^2\,\frac{r_\ell}{x'^{2}} \cos\theta\right\}\right).
\end{align}
where $N$ is the number of \lPBHs.

Since the final result depends on the spatial distribution of \lPBHs, encoded in  $f(\rvec_\ell - \mathbf{x}')$, we consider separately the cases of a uniform distribution, for which $f$ is spatially constant, and a non-uniform distribution, for which $f$ varies with position.

\subsubsection*{Uniform case}

We first consider a spatially homogeneous distribution of \lPBHs within the spike, a sphere of radius $L$. The corresponding single-particle distribution function is $
f(\rvec) \equiv \frac{1}{V}$, where $V=\frac{4}{3}\pi L^3$, and the 
total number of \lPBHs is $N=n V$, where $n$ denotes their uniform number density.

Performing the angular integrals in $\mathcal{J}$ and defining the dimensionless variable $z = (kGm^2 r_\ell)^{1/2}/x'$, one finds
\begin{align}
\mathcal{J}_u
= 4\pi\,(kGm^2 r_\ell)^{3/2}\,n
\int_0^\infty dz\,z^2\left[1 - z^2\sin(z^{-2})\right]
\approx 4.2\,(kGm^2 r_\ell)^{3/2}\,n \, ,
\end{align}
where the integral has been evaluated numerically. After azimuthal integration, the two-dimensional Fourier transform yields a Bessel function $J_0$, and introducing the dimensionless variable $s = \tau_{\rm c} k$ with the characteristic torque
\begin{align}
\tau_{\rm c} = 2.6\, Gm^2 r_\ell\, n^{2/3},
\end{align}
the probability density of the torque magnitude becomes
\begin{align}
\mathcal P(\tau_0|\rvec_\ell) = \frac{\tau_0}{\tau_{\rm c}^2}
\int_0^\infty ds\, s\, J_0\left(\frac{\tau_0}{\tau_{\rm c}}s\right)
e^{-s^{3/2}}\,.
\end{align}
Defining $\beta \equiv \frac{\tau_0}{\tau_{\rm c}}$,the distribution takes the scale-invariant form
\begin{align}
\mathcal{P}(\beta|\mathbf{r}_\ell)
= \beta\int_0^\infty ds\,s\,J_0(\beta s)\,e^{-s^{3/2}} \, .
\end{align}

\subsubsection*{Non-uniform case}

For a non-uniform \lPBH\ distribution, we adopt the power-law parametrization of Ref.~\cite{DelPopolo:1998in}, 
\begin{align}
f(\rvec) = \frac{1}{\mathcal V}\, r^{-p},
\qquad
\mathcal V = \frac{4\pi}{3-p}L^{3-p},
\qquad (0\le p<3),
\end{align}
where $\mathcal V = V\,\frac{3}{3-p}L^{-p}$ is the effective volume ensuring that $f(\rvec)$ is normalized to unity.

In analogy to the uniform case, we define the effective density
\begin{align}
\tilde n \equiv \frac{N}{\mathcal V}
= n(r)\,r^p,
\end{align}
where the second equality follows from $n(r) = Nf(r) = \tilde{n}\,r^{-p}$, and $\tilde{n}$ is a constant that characterizes the amplitude of the number-density profile (see \cref{sec:ell-ell} for the explicit profile used in the main text).

Proceeding as in the uniform case and working in the limit $x' \ll 1$, $\mathcal J$ acquires an explicit dependence on the inhomogeneity parameter $p$ through the 
modified $r_\ell$ scaling,
\begin{align}
\mathcal{J}_p
\approx 4.2\,(Gm^2 k)^{3/2} r_\ell^{(3-2p)/2} \tilde{n} \, ,
\end{align}
which reduces to $\mathcal{J}_u$ for $p = 0$ and $\tilde{n} = n$. The corresponding characteristic torque is
\begin{align}
\tilde{\tau}_{\rm c}
\approx 2.6\,Gm^2 r_\ell^{(3-2p)/3} \tilde{n}^{2/3} \, .
\end{align}
Despite the modified scaling of the characteristic torque, the probability density retains the same functional form as in the 
uniform case,
\begin{align}
\mathcal P(\tau_0|\rvec_\ell) \approx
\frac{\tau_0}{\tilde{\tau}_{\rm c}^{2}}
\int_0^\infty ds\, s\,
J_0\left(\frac{\tau_0}{\tilde{\tau}_{\rm c}}s\right)
e^{-s^{3/2}},
\end{align}
with $\mathcal P(\beta|\rvec_\ell)$ as in the uniform case, upon replacing $\tau_{\rm c} \to \tilde{\tau}_{\rm c}$.

\section{Primordial spectra of density perturbations and scales}
\label{app:CAMB}

\subsection*{Primordial curvature power spectra}
\label{app:subsec_PPS}

Let us now discuss the primordial curvature power spectra assumed in~\cref{sec:adiabatic}.
On large scales, observations of the cosmic microwave background (CMB) constrain the dimensionless primordial curvature power spectrum to a nearly scale-invariant power law,
\begin{equation}
\Delta_{\mathcal R, \rm CMB}^2(k)
=
A_s
\left(
\frac{k}{k_{\rm pivot}}
\right)^{n_s-1},
\end{equation}
with amplitude $A_s \simeq 2.1\times10^{-9}$, spectral index $n_s \simeq 0.965$, and pivot scale $k_{\rm pivot} = 0.05\,{\rm Mpc}^{-1}$~\cite{Planck:2018vyg}. This serves as our baseline CMB-only model.

For the enhanced spectra, required to accommodate PBH formation over an extended mass range, we rely on \textit{Model~(2)} proposed in~\cite{Franciolini:2022pav,Franciolini:2022tfm} and obtained from a 
reconstructed inflationary potential. This spectrum features a broad 
plateau on scales corresponding to PBH masses around and below $M_\odot$, while recovering the Planck-constrained behavior on large scales. It is built to produce \hPBHs~with masses $\simeq 1 M_\odot$ and a lighter component. To model a central \hPBH~of mass $M = 10^3\,M_\odot$, we also consider a shifted version of this spectrum, obtained by translating the plateau to the larger scales $k \sim k(10^3 M_\odot)$ associated with PBH formation at that mass.

\begin{figure}
    \centering
    \includegraphics[width=0.55\linewidth]{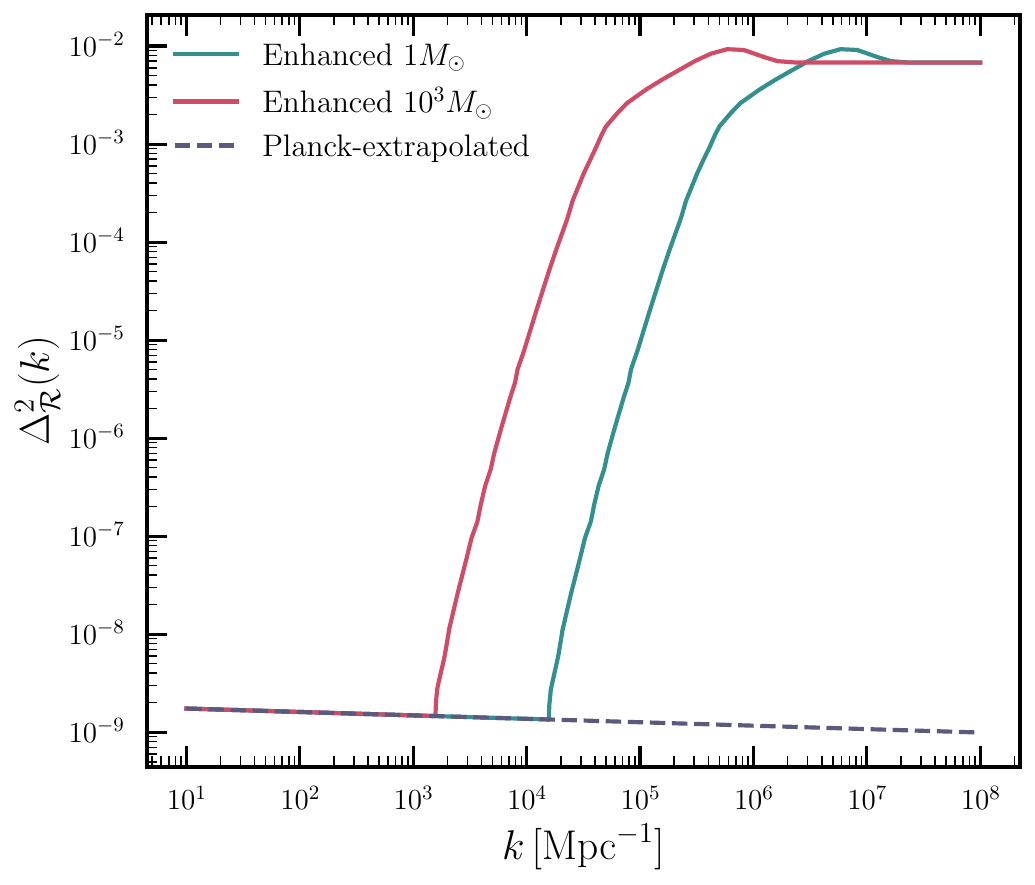}
    \caption{Primordial curvature power spectra considered in this work. The gray dashed line corresponds to the CMB-only baseline consistent with Planck observations~\cite{Planck:2018vyg}. The green solid line shows the enhanced spectrum of Refs.~\cite{Franciolini:2022pav,Franciolini:2022tfm} designed to produce PBHs around $M_\odot$. The dark red solid line is a shifted version of the same spectrum, used as a phenomenological benchmark for $M = 10^3\,M_\odot$. More details in the text.}
    \label{fig:PPS_evolution}
\end{figure}

As discussed in~\cref{sec:adiabatic}, our results are only weakly sensitive to the details of this small-scale enhancement; the shifted spectrum should be regarded as a phenomenological benchmark rather than a realistic prediction for PBH formation at $10^3 M_\odot$. The three primordial power spectra considered in this work are shown in Fig.~\ref{fig:PPS_evolution}. Note that both enhanced spectra recover the Planck-constrained behavior on large scales.

\subsection*{Evaluation of the variance of perturbations with \texttt{CAMB}} 
\label{app:subsec_CAMB}

\texttt{CAMB} \cite{Lewis:1999bs,camb_notes} is an Einstein-Boltzmann solver that can be used to evolve primordial curvature perturbations to redshift $z$. We use it to evolve the CMB-like baseline curvature spectrum $\Delta^2_{\mathcal{R},\rm CMB}(k)$ of \cref{app:subsec_PPS} to matter-radiation equality, extracting the CDM component $\Delta^2_{\rm CDM,CMB}(k,z_{\rm eq})$. From this, we derive the ratio
\begin{equation}
\label{eq:T_evol}
\mathcal{T}(k,z)\,= \frac{\Delta^2_{\rm CDM, CMB}(k,z)}{\Delta^2_{\mathcal R, \rm  CMB}(k)} \, , 
\end{equation}
which encapsulates both the linear evolution of perturbations and the mapping from primordial curvature perturbations to CDM density perturbations. Since linear perturbations evolve independently of the primordial spectrum, $\mathcal{T}(k,z)$ can be extracted once using the Planck-like baseline spectrum and then applied to any primordial spectrum. The CDM power spectrum at equality for each model of \cref{app:subsec_PPS} is therefore
\begin{equation}
\Delta^2_{\rm CDM}\left(k,z_{\rm eq}\right)
=
\mathcal{T}\left(k,z_{\rm eq}\right)\,
\Delta^2_{\mathcal R}(k).
\end{equation}

Since the computational cost limits \texttt{CAMB}'s applicability, for $k > 10^4\,{\rm Mpc}^{-1}$, where modes are deep inside the horizon at equality and $\mathcal{T}$ is well described by a power law, we extrapolate using a power-law fit to the \texttt{CAMB} output at the boundary. 
The resulting $\Delta^2_{{\rm CDM}}(k,z_{\rm eq})$ is then integrated to obtain $\sigma$
\begin{align}
\label{eq:tid_tensor_delta_complete}
    \sigma^2\equiv  \int d \ln k\; W^2(k)\,\Delta^2_{{\rm CDM}}(k,z_{\rm eq})\, ,
\end{align}
where the relevant scales are selected by convolving $\Delta^2_{{\rm CDM}}(k)$ with the window function
\begin{equation}
\label{eq:window}
W^2(k) = |W(k r_\mathrm{ta})|^2 - |W(k R_H(a_\mathrm{ta}))|^2\,.
\end{equation}
$W(X)$ is the Fourier transform of a spherical top-hat filter, $\rta$ is the turn-around radius and $R_H(a_\mathrm{ta})$ is the Hubble horizon at the time of turn-around.

\section{Crossing of \lPBH~orbits}
\label{app:orb_cross}

\begin{figure}[t!]
    \centering    \includegraphics[width=0.5\linewidth]{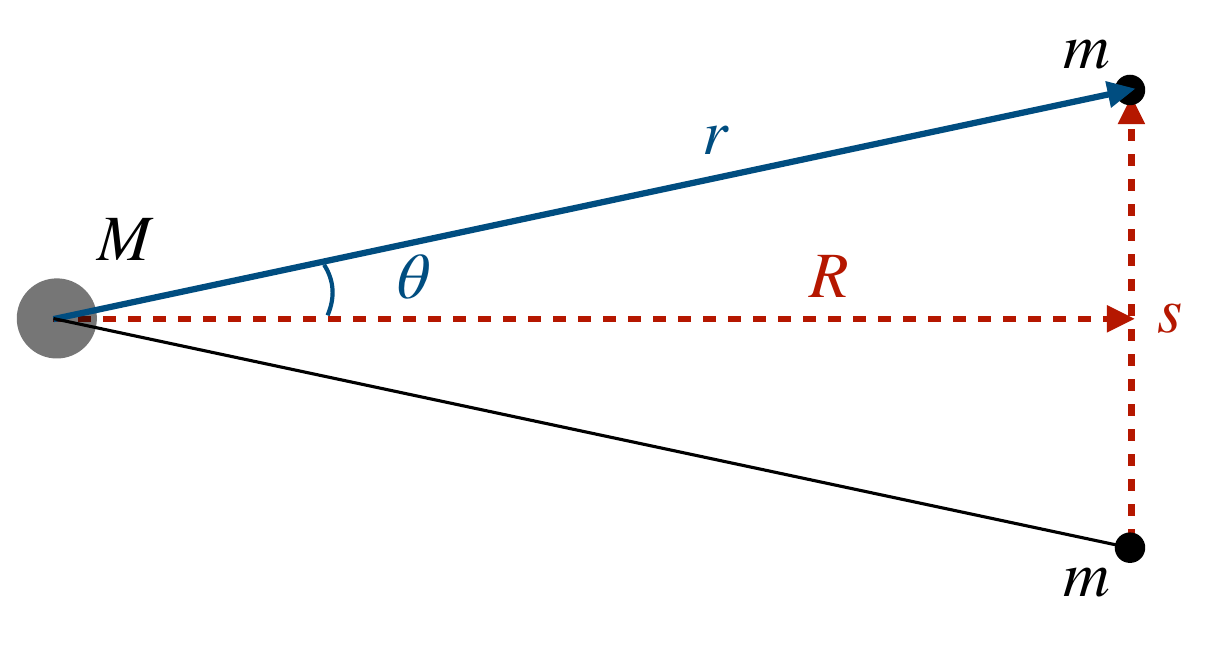}
    \caption{Setup for the study of the time of first crossing in 
    \cref{app:orb_cross}. The coordinates used in the full model are depicted in blue, those used in the tidal approximation in red. Each \lPBH~is at distance $r$ from the central \hPBH, at polar angle $\theta$ from the bisector of their separation $s$, which is assumed purely transversal. $R$ is the distance of the center of mass of the pair from the \hPBH.}
    \label{fig:crossing_setup}
\end{figure}

The computation of the light-light torques in \cref{sec:ell-ell} is valid as long as the \lPBHs~do not exchange positions during the infall toward the heavy one. To verify this, we study the evolution of two \lPBHs~in the \hPBH potential and compare the time at which their orbits first cross to the free-fall time. 
We take the \lPBHs~to be initially at rest, at equal distance $r$ from the 
\hPBH~and with purely transverse separation $x$. Any initial separation in the radial direction grows during infall, reducing mutual gravitational attraction; hence, a purely transversal separation minimizes the crossing time. 
Exploiting the symmetry of this configuration about the radial direction bisecting the pair, we can model the system in polar coordinates $(r,\theta)$, with 
$\theta$ the half-angle subtended by the pair at the \hPBH. The setup is shown in \cref{fig:crossing_setup}.

The Lagrangian for the motion of one particle at position 
$(r, \theta)$ reads
\begin{equation}
 \mathcal{L}=  \frac{m}{2} \left( \dot{r}^2 + r^2 \dot \theta^2 \right) + \frac{G Mm}{r} + \frac{G m^2}{4r \sin \theta} \, ,
\end{equation}
giving the coupled equations of motion for $\theta$ and $r$
\begin{equation}
\label{eq:crossing_full}
\left\{
\begin{aligned}
 \ddot r &= r \dot \theta^2 - \frac{G M}{r^2} - \frac{ G m}{4r^2 \sin \theta}\, ,\\
\ddot \theta &= - \frac{ 2 \dot r \dot \theta }{r}
 - \frac{ G m \cos \theta }{4 r^3 \sin^2 \theta} \,.
\end{aligned}
\right.
\end{equation}
Alternatively, we can treat this problem in the tidal approximation. In this limit, we have a purely radial two-body problem with an additional time-dependent harmonic potential due to the tidal field. In this case, the Lagrangian is
\begin{equation}
    \mathcal L = \frac{m}{2} \dot s ^2 + \dfrac{ 2 G m^2}{s} - \frac{m}{2} \kappa(t) s^2 \, ,
\end{equation}
where the elastic coupling is time-dependent and is given by $ \kappa(t)= GM/R(t)^3 $. The equation of motion along $s$ is
\begin{equation}
    \ddot s = - \frac{2 G m}{s^2} - \kappa(t) s \,, 
\end{equation}
where $R$ is the distance of the center of mass from the \hPBH. We obtain $\kappa(t)$ assuming that $R$ follows a free fall trajectory, solution of
$ \ddot R = - G M /R^2$.
The parameters $R$ and $s$ are related to the ones of the previous model by $R = r \cos \theta \simeq r$ and $s = 2 r \sin \theta \simeq 2 r \theta$, valid 
for $\theta \ll 1$.
It is convenient to use dimensionless variables and parameters
\begin{align}
    \rho &\equiv R/R_0\,, \quad  x  \equiv s/s_0\,, \\
    \tau &\equiv t/T_0 \,, \quad T_0= \left(\frac{R_0^3}{ GM}\right)^{\frac{1}{2}} \simeq 0.9 \, t_\mathrm{ff} \,,\\
    \alpha & \equiv \dfrac{m}{M} \left( \dfrac{R_0}{s_0}\right)^3 \simeq \dfrac{m}{M} \dfrac{1}{(2\theta_0)^3} \,,
\end{align}
where $R_0$ and $s_0$ are the initial conditions. In these variables, the equations of motion become
\begin{equation}
\label{eq:crossing_tid}
\left\{
\begin{aligned}
x'' &= - \frac{2\alpha}{x^{2}} - \frac{x}{\rho^{3}},\\
\rho'' &= - \frac{1}{\rho^{2}},
\end{aligned}
\right.
\end{equation}
with fixed initial conditions $x(0)=1, \, \dot x(0)=0 , \, \rho(0)=1$, and  $\dot \rho(0)=0 $. 

\begin{figure}
    \centering    \includegraphics[width=\linewidth]{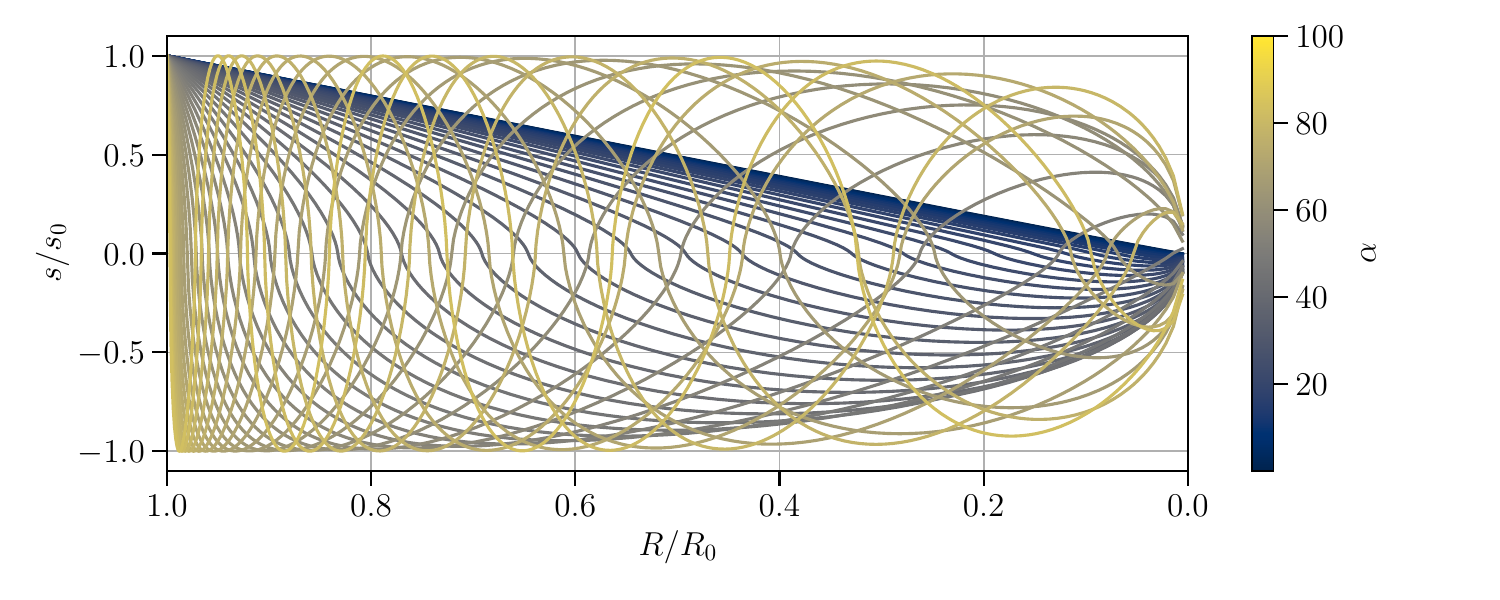}
    \caption{Solutions of equation \cref{eq:crossing_tid} for different values of the parameter $\alpha$. Two limiting behaviors exist: at large $\alpha$, mutual attraction dominates and the motion along $s$ is Keplerian, with repeated orbit crossings; at small $\alpha$, tidal force dominates and the trajectory tends to simple radial infall. In our problem $\alpha \simeq 0.1$, hence the latter regime applies.}
    \label{fig:orbits}
\end{figure}

This formulation makes it evident that there is only one degree of freedom for the system, parametrized by $\alpha$.
This parameter is given by the ratio of mutual gravity and tidal force and it controls the deviation of the system from the purely tidal case (no mutual force). The solutions to \cref{eq:crossing_tid} are shown in \cref{fig:orbits} for different values of $\alpha$.

\begin{figure}[h!]
    \centering    \includegraphics[width=\linewidth]{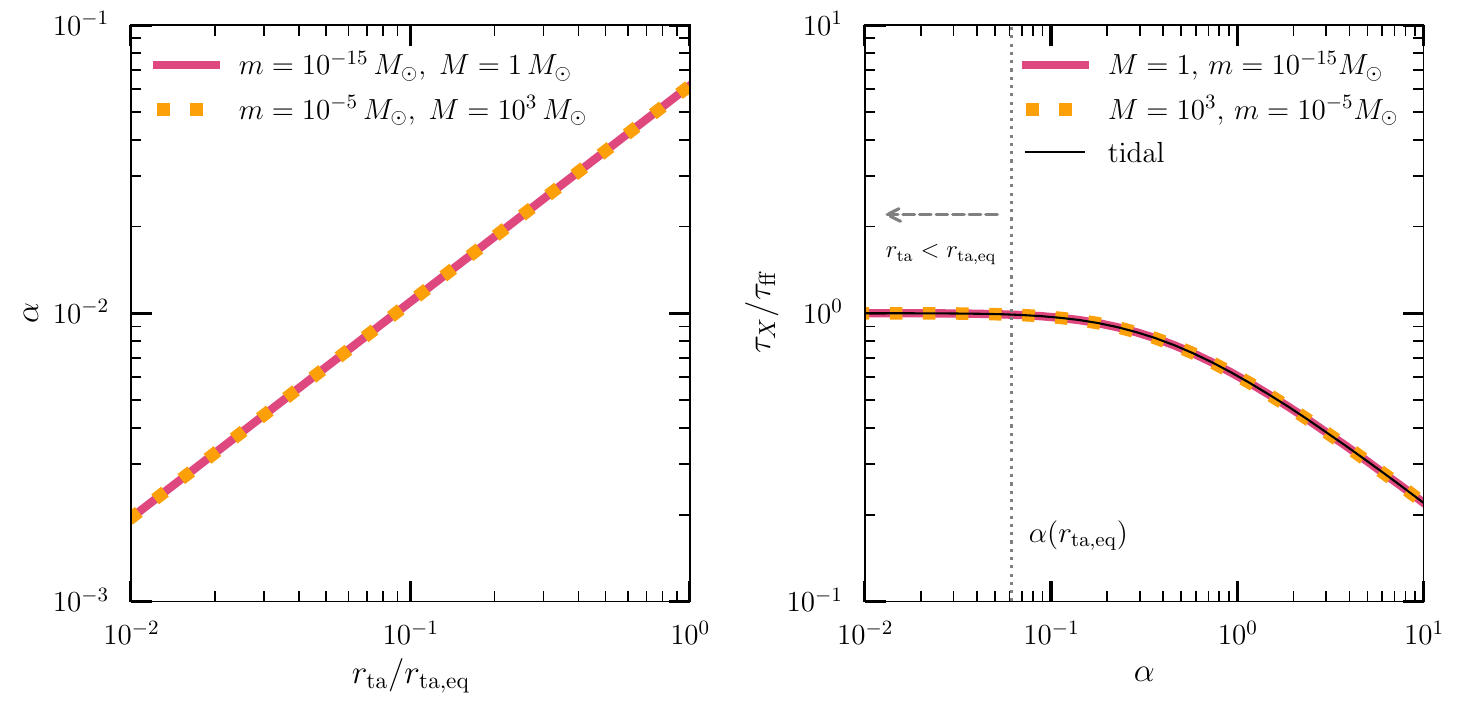}
    \caption{Time of the first crossing of orbits (in units of the free-fall time) as a function of the parameter $\alpha$. The vertical line indicates the expected value of $\alpha$ at the spike radius. Inner shells correspond to  smaller values. The dashed line is the prediction from model B, while the solid lines are obtained with model A for different mass ratios.}
    \label{fig:tau_crossing}
\end{figure}

Setting $s_0$ to the typical $\ell$--$\ell$ separation (\cref{eq:light_separation}) and $R_0$ to the turnaround radius, we can estimate $\alpha$ as a function of $\rta$ -- see the left panel of \cref{fig:tau_crossing}. Regardless of the mass values, $\alpha$ reaches its maximum at matter-radiation equality and in all shells $\alpha \lesssim 0.1$.
Hence, the tidal force dominates and the motion is initially very close to a pure radial trajectory. In the right panel of \cref{fig:tau_crossing} we show the time of first crossing $\tau_X$ in units of the free-fall time, as a function of $\alpha$. The black line is obtained varying $\alpha$ and solving \cref{eq:crossing_tid}, while the magenta and orange lines are the crossing times obtained from \cref{eq:crossing_full} varying the initial separation $x$.
We find $\tau_X/ \tau_\mathrm{ff} \gtrsim 0.97 $ for $\alpha \lesssim 0.1$: for typical nearest-neighbor separations within the shells, the time elapsed up to the crossing is almost as large as the total free-fall time. Therefore, we can  assume that the spatial ordering of \lPBHs is preserved during their infall.

\section{Simulation setup}
\label{app:simulation}

We use \texttt{GADGET-4}~\cite{Springel:2020plp} to perform $N$-body simulations of the infall of \lPBHs~toward a central \hPBH. The setup is designed to capture the dynamics of the \lPBHs under the combined influence of the central heavy seed and the stochastic torques generated by \lPBH-\lPBH~interactions. Since the simulations cannot resolve the dynamics down to the Schwarzschild radius of the \hPBH, they cannot be used to study direct capture.
The simulation pipeline and the adopted parameter values are publicly available at \url{https://github.com/atolino/Light_PBH_spikes}. The central \hPBH is modeled as a fixed external Hernquist 
potential~\cite{Hernquist:1990be}, which reduces to a Newtonian point mass on scales larger than the scale radius $r_H$. Both the scale radius and the gravitational softening length are set to the minimum values consistent with numerical stability and reliable resolution of the minimum approach distance between particles.
Force accuracy and time-stepping tolerances are calibrated to reliably capture the dynamical evolution, while maintaining a manageable computational cost.

The simulated region is a sphere of radius $L = r_{\rm sp,eq} \simeq 10^{-2}\,\mathrm{pc}$, chosen to match the spike radius at equality for the $M = M_\odot$ benchmark. The \lPBH~number density within this volume is $n = 3N_{\rm sim}/(4\pi L^3)$, where $N_{\rm sim}$ ranges from $10^4$ to $10^5$ depending on the mass ratio $m/M$ and the required dynamic range. Each simulation is then evolved up to the free-fall collapse time of the initial configuration, $t_{\rm ff} \sim 
\sqrt{3\pi/(32 G \bar\rho_\ell)}$, where $\bar\rho_\ell = m\,n$ is the mean \lPBH~mass density.

\section{Torque evolution during infall}
\label{app:scale_inv}

Here, we explore the evolution of the torque due to \lPBH-\lPBH~interactions as the \lPBHs~fall toward the central \hPBH, in order to justify the constant-torque approximation used in \cref{eq:ell_of_tau}. 
We consider a reference \lPBH~at position $\mathbf{r}$ with respect to the \hPBH, and a neighboring \lPBH~at $\mathbf{r}^\prime = \hat{\mathbf{r}}\,(r + s) + \mathbf{\xi}$, where $s$ is the radial separation and $\mathbf{\xi} \perp \hat{\mathbf{r}}$ is the transverse separation. The torque on the first \lPBH~due to the second is
\begin{align}
    \boldsymbol{\tau} &= - G m^2 \frac{\mathbf{r} \times \mathbf{r}^\prime}{\left| \mathbf{r} - \mathbf{r}^\prime\right|^3} = -G m^2 \frac{\mathbf{r}\times \mathbf{\xi}}{\left(s^2 + \xi^2\right)^{3/2}}\, ,
\end{align}
where we used $\mathbf{r}\times\hat{\mathbf{r}} = 0$ in the last step.

In the limit $s,\,\xi \ll r$, the tidal acceleration  due to the field of the central \hPBH\ is given by~\cite{Pinochet:2022ooy}:
\begin{equation}
    \ddot{s} = 2 G M \frac{s}{r^3}\, , \,\,\,\ddot{\xi} = -GM\frac{\xi}{r^3}\,.
\end{equation}
These equations admit a power-law solution as a function of $r$, giving the evolution of the separations as:
\begin{align}
\begin{split}
    s = s_i \left(\frac{r}{r_i}\right)^{-1/2} \, , \,\,\,
    \xi = \xi_i \left(\frac{r}{r_i}\right)\,,
    \end{split}
\end{align}
for infall from some initial radius $r_i$.
This reflects the well-known \textit{spaghettification}: radial separations grow as $r^{-1/2}$ while transverse separations shrink as $r$, with the product $\xi s^2 \propto \mathrm{constant}$, conserving the volume element of an initially spherically distributed collection of objects.

Writing the initial separation as $\delta r_i = \sqrt{s_i^2 + \xi_i^2}$ 
and decomposing $(\xi_i, s_i) = \delta r_i(\sin\vartheta, \cos\vartheta)$, the torque magnitude becomes
\begin{align}
\label{eq:tau_theta}
    \tau
    = \frac{Gm^2 r_i}{\delta r_i^2}
    \frac{\tilde{r}^2\sin\vartheta}
    {\left(\sin^2\vartheta\,\tilde{r}^{2} 
    + \cos^2\vartheta\,\tilde{r}^{-1}\right)^{3/2}},
\end{align}
where $\tilde{r} \equiv r/r_i$. The angle $\vartheta \in [0, \pi]$ parametrizes the orientation of the neighbor: $\vartheta = 0$ and $\vartheta = \pi$ correspond to purely radial separation and $\vartheta = \pi/2$ to purely transverse separation. For an isotropic distribution, 
$\cos\vartheta$ is uniformly distributed on $[0,1]$.

\begin{figure}
    \centering
    \includegraphics[width=0.5\linewidth]{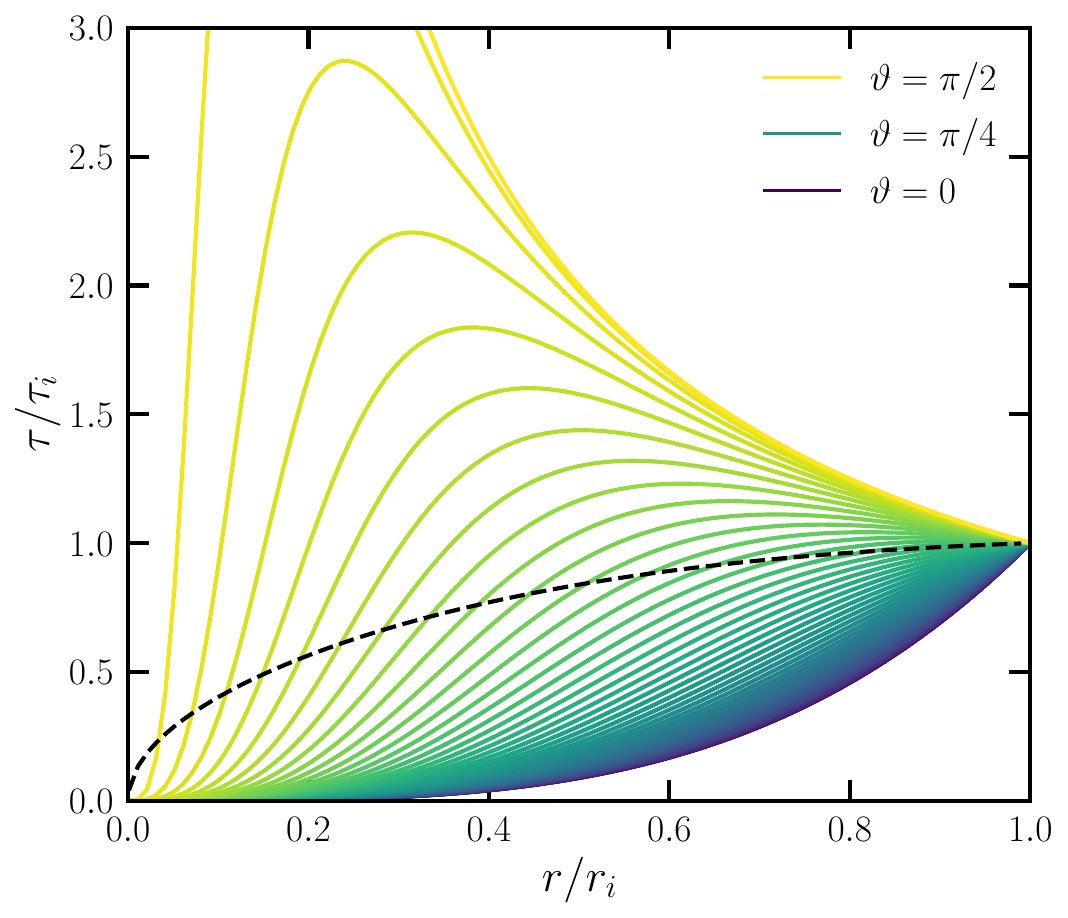}
    \caption{Evolution of the torque on a \lPBH~due to a \textit{single} neighboring \lPBH, as a function of the normalized radial distance 
    $\tilde{r} = r/r_i$, for several values of the orientation angle 
    $\vartheta$ (colored lines, \cref{eq:tau_theta}). The angle $\vartheta$ defines the relative orientation of the neighboring \lPBH; $\vartheta = 0$, corresponds to purely radial separation and 
    $\vartheta = \pi/2$ to purely transverse separation. The evolution of $\tau$ is symmetric about $\vartheta = \pi/2$, so we do not show curves for $\vartheta > \pi/2$. The dashed black line shows the average over orientations
    (\cref{eq:tau_average}).}
    \label{fig:torque_evolution}
\end{figure}

The evolution of $\tau$ for different values of $\vartheta$ is shown in  \cref{fig:torque_evolution}. For (almost) purely radial configurations
($\vartheta \sim 0$), the torque decreases rapidly as the radial 
separation of the two \lPBHs~grows during infall. For more transverse configurations, the torque initially grows as $\xi$ shrinks, before eventually decreasing as the growth in radial separation $s$ dominates. In all cases, the torque varies by at most a factor of a few throughout most of the infall, with the largest  variations occurring only near $r \to 0$ where the infall velocity is maximal and the contribution to the accumulated angular momentum is negligible.
Averaging \cref{eq:tau_theta} over $\cos\vartheta$, we find that the mean torque is
\begin{align}
    \langle\tau\rangle 
    = \frac{Gm^2 r_i}{\delta r_i^2}\,I(\tilde{r})\, ,
\end{align}
where
\begin{align}
\label{eq:tau_average}
    I(\tilde{r}) 
    = \frac{4}{\pi}\frac{\tilde{r}^2}{\tilde{r}^3-1}
    \left[K\!\left(1-\tilde{r}^{-3}\right) 
    - E\!\left(1-\tilde{r}^{-3}\right)\right],
\end{align}
with $K(m)$ and $E(m)$ the complete elliptic integrals of the first and second kind with parameter $m$, and normalized such that $I(\tilde{r}) \to 1$ as  $\tilde{r} \to 1$. The function $I(\tilde{r})$ is shown as a dashed  line in \cref{fig:torque_evolution}.
With this definition, we can expect that on average the evolution of the torque is determined by $I(\tilde{r})$, i.e. $\tau(r) = \tau_i \times I(r/r_i)$ .

As seen in \cref{fig:torque_evolution}, the evolution of the torque depends on the relative orientation of the nearest neighbor, or, more generally, the relative configuration of all nearby \lPBHs. $I(\tilde{r})$ remains of order unity throughout most of the infall and declines only as $r \to 0$. This confirms that the torque is approximately constant during the bulk of the collapse, justifying the constant-torque approximation in~\cref{eq:ell_of_tau}, and that the dominant contribution to the accumulated angular momentum is acquired at large radii near turn-around.

\clearpage

\bibliographystyle{utphys}
\bibliography{main}  
\end{document}